\documentclass[12pt]{iopart}

\usepackage{iopams}  
\usepackage{bm}
\usepackage{multirow}
\usepackage{graphicx} 
\usepackage{subcaption}
\usepackage{threeparttable} 
\usepackage[colorlinks=true,urlcolor=blue]{hyperref}  
\usepackage{algorithm} 
\usepackage{algpseudocode} 
\usepackage{booktabs}
\usepackage{longtable}
\usepackage{xcolor}

\definecolor{darkred}{RGB}{139, 0, 0}
\definecolor{darkblue}{RGB}{0, 100, 170}

\newcommand{\WC}[1]{\textcolor{gray}{#1}}
\newcommand{\WP}[1]{\textcolor{darkblue}{#1}}
\newcommand{\MWP}[1]{\textcolor{darkred}{#1}}

\begin{document}

\title[ChatBCI:\@ A P300 Speller BCI Leveraging ChatGPT for Sentence Composition]{ChatBCI: A P300 Speller BCI Leveraging\\ Large Language Models for Improved Sentence Composition in Realistic Scenarios}

\author{Jiazhen Hong$^+$, Weinan Wang$^+$, and Laleh Najafizadeh}
\address{Integrated Systems and NeuroImaging Laboratory \\Department of Electrical and Computer Engineering, Rutgers University \\Piscataway, NJ 08854, USA \\
($^+$Equal Contributions)\\
\footnote[0]{\hrule \ \\ A demo video can be found \href{https://www.youtube.com/watch?v=UxSWutNDECM}{here}.}\\
}
\vspace{-6mm}
%%%%%%%%%%%%%%%%%%
%%%% ABSTRACT %%%%
%%%%%%%%%%%%%%%%%%
\begin{abstract} P300 speller brain computer interfaces (BCIs) allow users to compose sentences by selecting target keys on a graphical user interface (GUI) through the detection of P300 component in their electroencephalogram (EEG) signals following visual stimuli. Most existing P300 speller BCIs require users to spell all or the first few initial letters of the intended word, letter by letter. Consequently, a large number of keystrokes are required to write a sentence, which can be time consuming, increasing user’s cognitive load and fatigue. Therefore, there is a need for more efficient and user-friendly methods for faster, and practical sentence composition.

In this work, we introduce ChatBCI, a P300 speller BCI that leverages the zero-shot learning capabilities of large language models (LLMs) to suggest words from user-spelled initial letters or predict the subsequent word(s), reducing keystrokes and accelerating sentence composition. ChatBCI retrieves word suggestions through remote queries to the GPT-3.5 API. A new GUI, displaying GPT-3.5 word suggestions as extra keys is designed. Stepwise linear discriminant analysis (SWLDA) is used for the P300 classification. 

Seven subjects completed two online spelling tasks: 1) copy-spelling a self-composed sentence using ChatBCI, and 2) improvising a sentence using ChatBCI’s word suggestions. Results demonstrate that in Task 1, on average, ChatBCI outperforms letter-by-letter BCI spellers, reducing time and keystrokes by $62.14\%$ and $53.22\%$, respectively, and increasing information transfer rate by $198.96\%$. In Task 2, ChatBCI achieves $80.68\%$ keystroke savings and a record $8.53$ characters/min for typing speed. 

Overall, ChatBCI by employing remote LLM queries enhances sentence composition in realistic scenarios, significantly outperforming traditional spellers without requiring local model training or storage. ChatBCI’s (multi-) word predictions, combined with its new GUI, pave the way for developing next-generation speller BCIs that are efficient and effective for real-time communication, especially for users with communication and motor disabilities.

\end{abstract}

\vspace{-2mm}
\normalsize
\maketitle

%%%%%%%%%%%%%%%%%%%%%%%%%%%%%%%%%%%%%
%%                                 %%
%% ------ 1. INTRODUCTION -------  %%
%%                                 %%
%%%%%%%%%%%%%%%%%%%%%%%%%%%%%%%%%%%%%
\section{Introduction}\label{sec:intro}

Event-related potentials (ERPs), which appear in electroencephalogram (EEG) signals in response to an external stimulus, have been long utilized in developing P300 speller brain-computer interfaces (BCIs) \cite{rezeika2018brain, cecotti2011spelling}. These speller BCIs particularly find potential applications in aiding individuals with communication and motor disabilities, such as those with amyotrophic lateral sclerosis (ALS), to express their thoughts and needs. 

The P300 component, a positive peak in ERP that occurs approximately $300$ ms after the onset of stimulus onset, has been considered a reliable signal in speller BCIs \cite{farwell1988talking, wolpaw2002brain, krusienski2008toward, comaniciu2018enabling}. In a classical P300 speller BCI, keys in the graphical user interface (GUI) are arranged in a $6 \times 6$ matrix \cite{farwell1988talking}, with each row and column in the matrix flashing repeatedly in a random sequence to generate visual stimulations. When the user focuses on a particular key, the flashing of the row and the column containing the desired key is recognized as a deviant stimulus, evoking a P300 response similar to the oddball paradigm \cite{fabiani1987definition, karamzadeh2013capturing}. Due to the low signal-to-noise ratio (SNR) of single-trial EEG signals, generally multiple repetitions are required, to confirm a selection. 

Over the past few decades, efforts have been made to improve BCI spellers by enhancing the speed and accuracy of making P300-based key selections. These include optimizing the GUI \cite{1214695}, enhancing flashing patterns \cite{jin2010new, townsend2010novel}, and adjusting inter-stimulus intervals (ISIs) \cite{mcfarland2011p300, lu2013effects, han2022effects}. These efforts however, did not target improving the spelling efficiency, as selections were mostly made letter by letter, limiting the usability of speller BCIs for composing long messages due to the fatigue and cognitive load caused by the prolonged focus required to select numerous characters from the keyboard.

Incorporating natural language processing (NLP) in P300 speller BCIs has been suggested recently to take advantage of NLP's ability to predict and suggest words based on partial input. By learning the linguistic properties of the natural language from a training corpus, statistical language models such as N-Gram or probabilistic automata have been used to estimate the probabilities for the upcoming letter or word, given the partially completed text composed by the user to enhance the spelling process.  

Table~\ref{tab:1} summarizes recent P300-based speller BCI studies that have incorporated NLP and evaluated the performance in online spelling experiments. These studies are categorized into two groups based on their objectives: predicting the probability of the upcoming letter \cite{speier2015incorporating, mainsah2014utilizing, ulacs2013incorporation, Chen2022Language}, or automatically completing the word after typing its initial few letters \cite{speier2018improving, chandravadia2022comparing, kaufmann2012spelling, ryan2010predictive, Akram2013novel}.  The studies in \cite{speier2011natural,parthasarathy2020optimizing, parthasarathy2024high} also incorporated NLP techniques with P300 speller BCIs, however, their focus was limited to offline or simulation experiments, and hence are not included in the table. When looking at the performance results (e.g., selection accuracy, average time to complete, and information transfer rate (ITR)) with and without the use of NLP, it is evident that incorporating NLP improves the performance of P300 speller BCIs. However, most of these implementations require additional time, software and hardware budgets to create a corpus; implement, train and update language models; and run model inferences, all of which increase the deployment costs of the BCIs. Additionally, these implementations are unable to predict the next word based on the context of preceding words to further improve their spelling efficiency. Moreover, the evaluation of these speller BCIs has mostly been limited to word spelling tasks (i.e., spelling a set of randomly selected words), which rarely occur in real-life scenarios. Sentence spelling tasks, by contrast, are more representative of real-life conversations and better suited for evaluating the performance of speller BCIs in practical applications.

\begin{table}
    \centering
    \caption{\small Summary of recent P300 speller BCI studies utilizing language models. Results averaged across all subjects are reported. ITR: information transfer rate calculated based on the number of \textit{selections}. ITR$^{*}$: information transfer rate calculated based on the number of \textit{characters}.}\label{tab:1}
    \resizebox{\textwidth}{!}{%
    \begin{threeparttable}
\begin{tabular}{l|ccl|p{5.2cm}|lcc}
\hline
\textbf{NLP Objective} & \textbf{Reference} & \textbf{Language} & \textbf{Language Model} & \textbf{\begin{tabular}[c]{@{}l@{}} Spelling Task (Online) \\ (\# of total characters)\end{tabular}} & \textbf{\begin{tabular}[c]{@{}l@{}}Performance \\ Metric\end{tabular}} & \textbf{\begin{tabular}[c]{@{}c@{}}Performance \\ Without NLP\end{tabular}} & \textbf{\begin{tabular}[c]{@{}c@{}}Performance \\ With NLP\end{tabular}} \\ \hline
\multirow{5}{*}{\begin{tabular}[c]{@{}l@{}}Letter probability\\ estimation\end{tabular}} & \cite{speier2015incorporating} & English & Trigram & one phrase (-) & ITR (bits/min) & N/A & 30.69 \\
 & \cite{speier2015incorporating} & English & Probabilistic Automata & one phrase (-) & ITR (bits/min) & N/A & 37.31 \\
 & \cite{mainsah2014utilizing} & English & Bigram & five 6-letter words, one 6-digit number (36) & ITR (bits/min) & 29.55 & 33.15 \\
 & \cite{ulacs2013incorporation} & Turkish & Trigram & four words (26) & ITR (bits/min) & 15.38 & 23.88 \\
 & \cite{Chen2022Language} & English & N-gram & twelve 6-letter words (72) & Selection Accuracy & 68.1\% & 79.6\% \\ \hline
\multirow{5}{*}{\begin{tabular}[c]{@{}l@{}}Word\\ completion\end{tabular}} & \cite{speier2018improving} & English & Probabilistic Automata & ten words (-) & ITR\tnote{*} (bits/min) & 53.89 & 59.39 \\
 & \cite{chandravadia2022comparing} & English & Probabilistic Automata & three 10-letter words (30) & ITR\tnote{*} (bits/min) & N/A & 72.11 \\
 & \cite{kaufmann2012spelling} & German & Dictionary & one 9-word sentence (45) & ITR\tnote{*} (bits/min) & 12.0 & 20.6 \\
 & \cite{ryan2010predictive} & English & Commercial Product (WordQ2) & one 10-word sentence (58) & Time to Complete (min) & 20.20 & 12.43 \\
 & \cite{Akram2013novel} & English & Dictionary & ten words (-) & Word Typing Time (min) & 2.9 & 1.66 \\ \hline
\end{tabular}%
      \end{threeparttable}
   }
 \vspace{-5mm}
\end{table}

Recently, the development of large language models (LLMs) has enhanced the ability to perform various NLP tasks without specific customization. Transformer-based LLMs, when scaled and trained on large datasets, display new capabilities \cite{Wei2022}. One example is zero-shot learning, in which LLMs can follow instructions described in natural languages (known as ``prompts'') to perform versatile and highly-specialized NLP tasks such as question answering, translation, and text generation \cite{Brown2020}. This allows for rapid, low-cost implementation of complex NLP tasks via remote queries to a general LLM, eliminating the need for local model deployment. A widely recognized implementation of such LLM is ChatGPT \cite{ChatGPT} (e.g., GPT-3.5 \cite{Brown2020}) with a publicly accessible application programming interface (API).

Motivated by the advanced capabilities of LLMs, here, we present ChatBCI, a P300 speller BCI that offers predictive spelling features entirely through remote GPT-3.5 queries. Leveraging the zero-shot learning capabilities of LLMs, we use a prompt template to request candidate words from the GPT-3.5 API for completing the partial sentence composed by the user, thereby, enhancing the entire spelling process. To facilitate this for the user, a new GUI is designed for the ChatBCI. As will be demonstrated, the integration of GPT-3.5 into ChatBCI not only enables word completion, but also allows for predicting the next word or multiple words in a sentence based on the context of the preceding words, consequently, minimizing the number of required selections and significantly accelerating the sentence composition process.\bigskip
\\
The key contributions and novelties of this paper are as follows:
\begin{itemize}
    \item[-] We introduce ChatBCI, the first P300 speller BCI to incorporate LLMs (here GPT-3.5), providing word completion and (multi-)word prediction capabilities to improve typing speed, performance, and user experience.\medskip
    \item[-] We present a new keyboard GUI that displays candidate words suggested by the LLM in addition to the basic characters in traditional speller BCIs, allowing users to type efficiently and customize their experience based on personal preferences. \medskip 
    \item[-] Using a carefully designed prompt template, we fully utilize the NLP capability of GPT-3.5 to offer intelligent word suggestions. While preserving the basic word completion function achieved by previous statistical NLP models, GPT 3.5 further enables direct prediction of the upcoming word(s) in a sentence, for additional keystroke savings. GPT-3.5 intelligently switches between word completion and prediction based on the completeness level of the last word in the partially spelled sentence, without requiring any modifications to the prompt template or the layout of the word suggestion panels in the GUI. In contrast, achieving the same level of functionality with traditional statistical NLP models would have required a far more complex system.  \medskip
    \item[-] In addition to the online copy-spelling writing task used in previous studies for evaluating speller performance, we conduct a new online experiment asking subjects to improvise sentences using ChatBCI, to evaluate the speller's efficiency in a more naturalistic and practical scenario, better reflecting real-world communication needs. \medskip
    \item[-] We propose to use keystroke analysis to quantify the predictive typing capabilities of speller BCIs and introduce a new metric, keystroke savings deficit ratio (KS-DR), to assess how close a speller BCI with predictive capabilities is to achieve its theoretical maximum keystroke savings.
\end{itemize}
\medskip

Overall, through integration of LLMs, we foresee great potentials in ChatBCI for significantly improving the efficiency and user experience of P300 spellers, without introducing burdens of training, storing, or running language models locally. ChatBCI therefore, represents a new generation of P300 speller BCIs that enhances spelling experiences for the users, offering potentials for developing real-time communication tools in assistive technologies.

The rest of the paper is organized as follows. Section 2 describes the details of the proposed ChatBCI including the GUI, LLM integration, and the classification algorithm, as well as the metrics used for evaluating the performance of the speller. Sections 3 outlines the experimental procedures, including the two online spelling tasks (copy-spell and improvisation). The results are presented in Section 4 , followed by discussions in Section 5. Finally, the paper is concluded in Section 6.

%%%%%%%%%%%%%%%%%%%%%%%%%%%%%%%%%%%%%
%%                                 %%
%% ------  Methods  ------  %%
%%                                 %%
%%%%%%%%%%%%%%%%%%%%%%%%%%%%%%%%%%%%%

\section{Methods}\label{sec:method}

\begin{figure}
  \centering 
  \includegraphics[width=\linewidth]{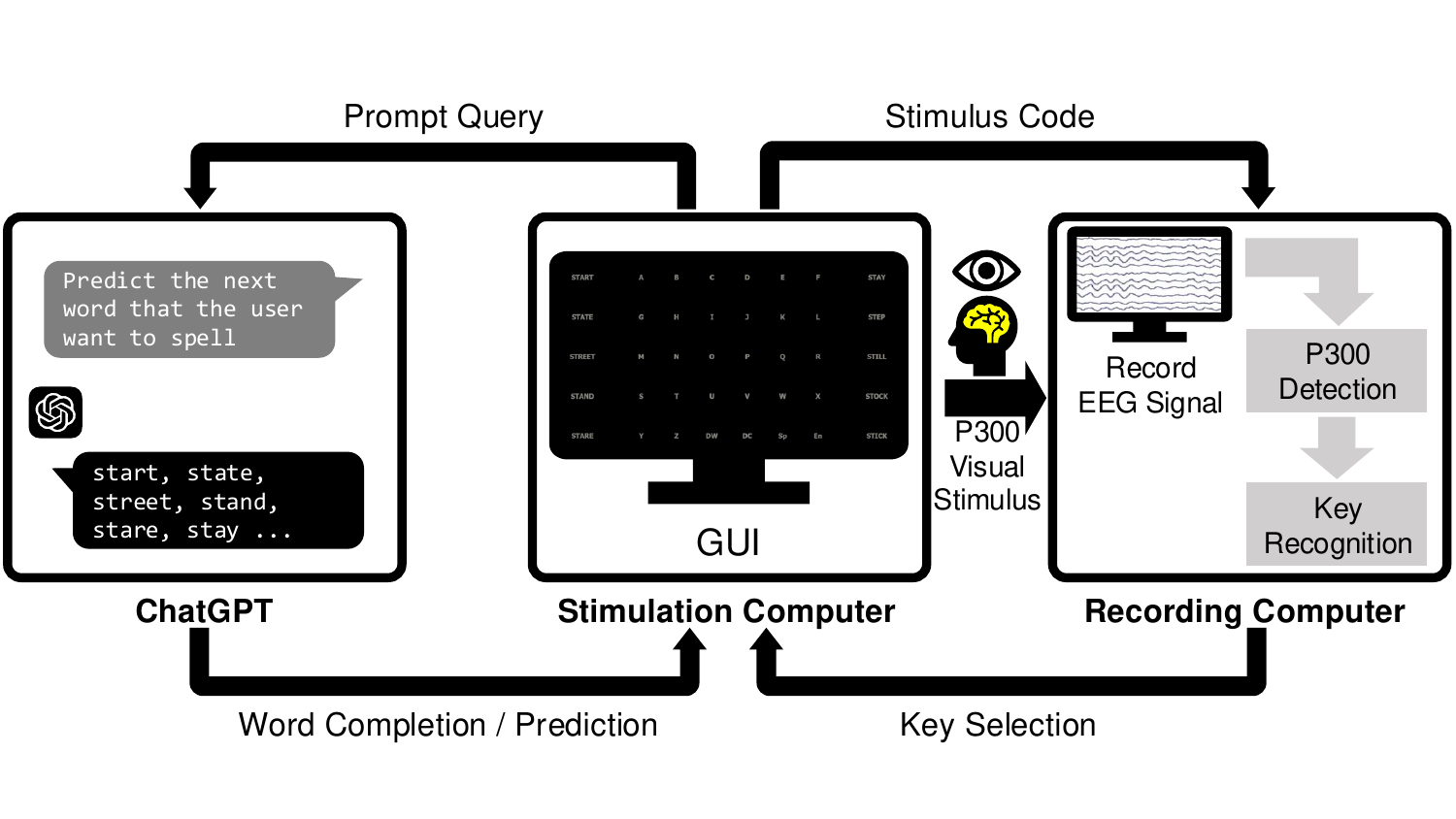} 
  \caption{\small Overview of the proposed ChatBCI. The new keyboard GUI, integrated with remote ChatGPT query, is displayed on the stimulation computer. The recording computer records and processes the acquired EEG signals in real time for P300 detection and key recognition. Key selections detected by the recording computer update the relevant panels in the GUI, displaying the spelled sentence and suggested words.}
  \label{fig:2}
\end{figure}

Figure \ref{fig:2} provides an overview of the proposed ChatBCI, which consists of a stimulation computer, a recording computer, and an EEG recoding system. The stimulation computer displays the LLM-integrated GUI and transmits stimulus codes, indicating the sequence of row and column flashes on the GUI, to the recording computer. The recording computer records and processes the acquired EEG signals, and infers the user's selected key by matching the detected P300 ERP with the flashing pattern. The selected key is then sent back to the stimulation computer to update the GUI, and remote LLM queries are conducted to retrieve and update word suggestions displayed on the GUI based on the newly-selected key. In what follows, we describe the GUI, explain how LLM is integrated with the BCI speller, and discuss how key selections are determined via detection of P300 ERP.

%%%%%%%%%%%%%%%%%%%%%%%%%%%%%%
\subsection{Graphical User Interface (GUI) of the ChatBCI}
%%%%%%%%%%%%%%%%%%%%%%%%%%%%%%

\begin{figure}
    \centering 
    \includegraphics[width=0.9\textwidth]{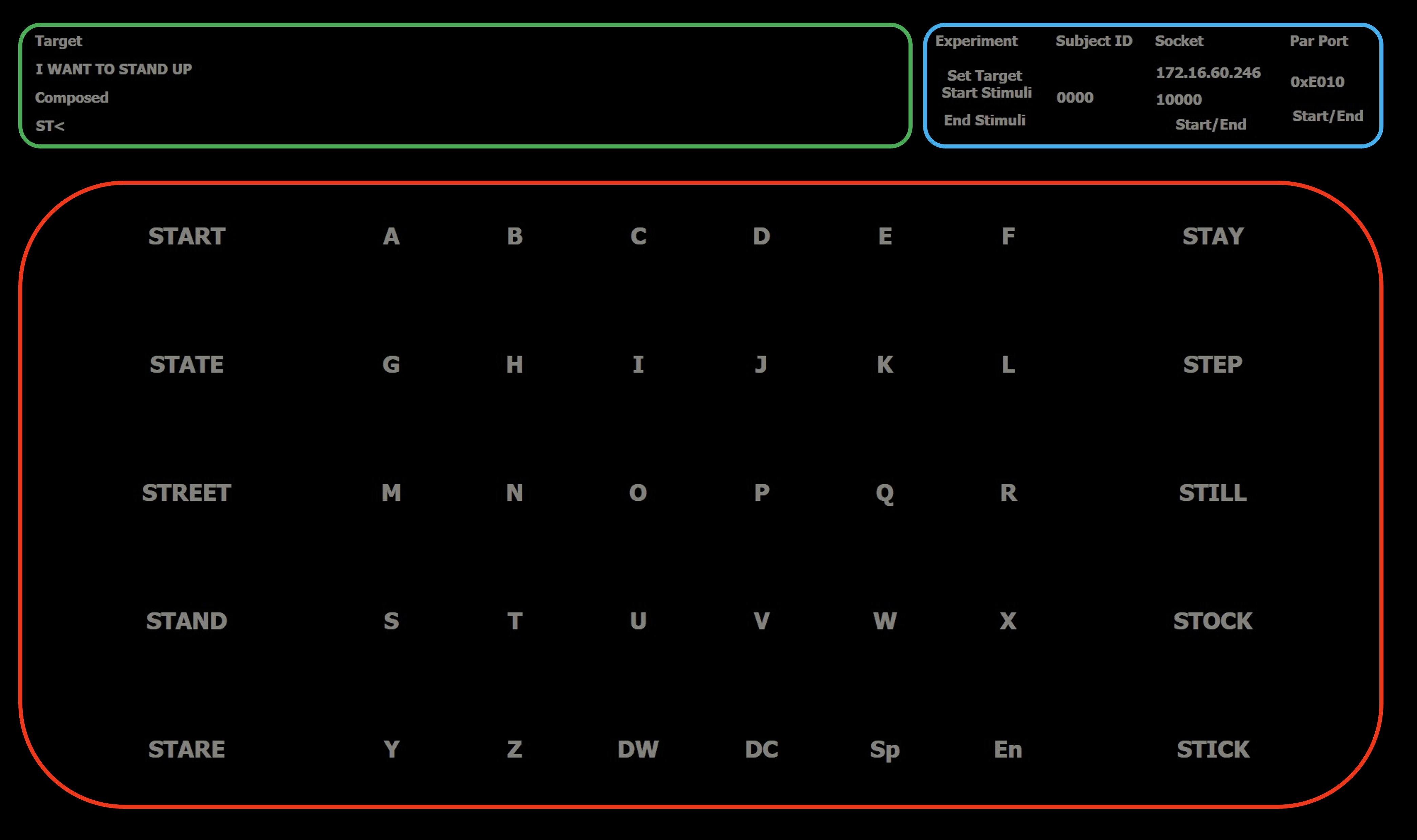} 
    \caption{\small The new keyboard GUI developed for ChatBCI. \textbf{Top-right}: the sentence panel (shown within the green box) displays two sentences: the target sentence (e.g., used in the copy-spell task (Task 1)), and the sentence as composed by the user in real-time. \textbf{Top-left}: the experiment panel (shown within the blue box) presents information about the experiment. \textbf{Bottom}: the keyboard panel (shown within the red box) includes character keys, function keys, and $5$ slots on the left and right sides, to display the $10$ candidates suggested by GPT.}
    \label{fig:3}
  \end{figure}

Figure~\ref{fig:3} illustrates the GUI developed for the proposed ChatBCI. The GUI has three panels. The main body of the GUI, the keyboard panel (shown within the red box), is a \(5 \times 8\) matrix of keys, and includes $26$ alphabet keys for letters A to Z, four function keys, and two $1\times5$ columns on the left and right sides to display the $10$ suggested candidates that are provided dynamically through GPT-3.5 queries. The function keys provide basic editing functions for the user when composing sentences: `DW' (delete word) removes the last word, whether complete or incomplete, from the current sentence; `DC' (delete character) removes the last character; `Sp' inserts a space; and `En' indicates the completion of an entry. In the sentence panel (shown within the green box), the ``Target'' section displays the target sentence for copy-spelling experiments, and the ``Composed'' section shows the sentence composed by the user using ChatBCI in real-time. For better visual visibility, in the ``Composed'' section, all letters are shown in upper case, and all space characters are shown as dashes (``-''). The experiment panel (shown within the blue box) gathers information about the experiment. 

\begin{figure}
	\centering 
	\includegraphics[width=\textwidth]{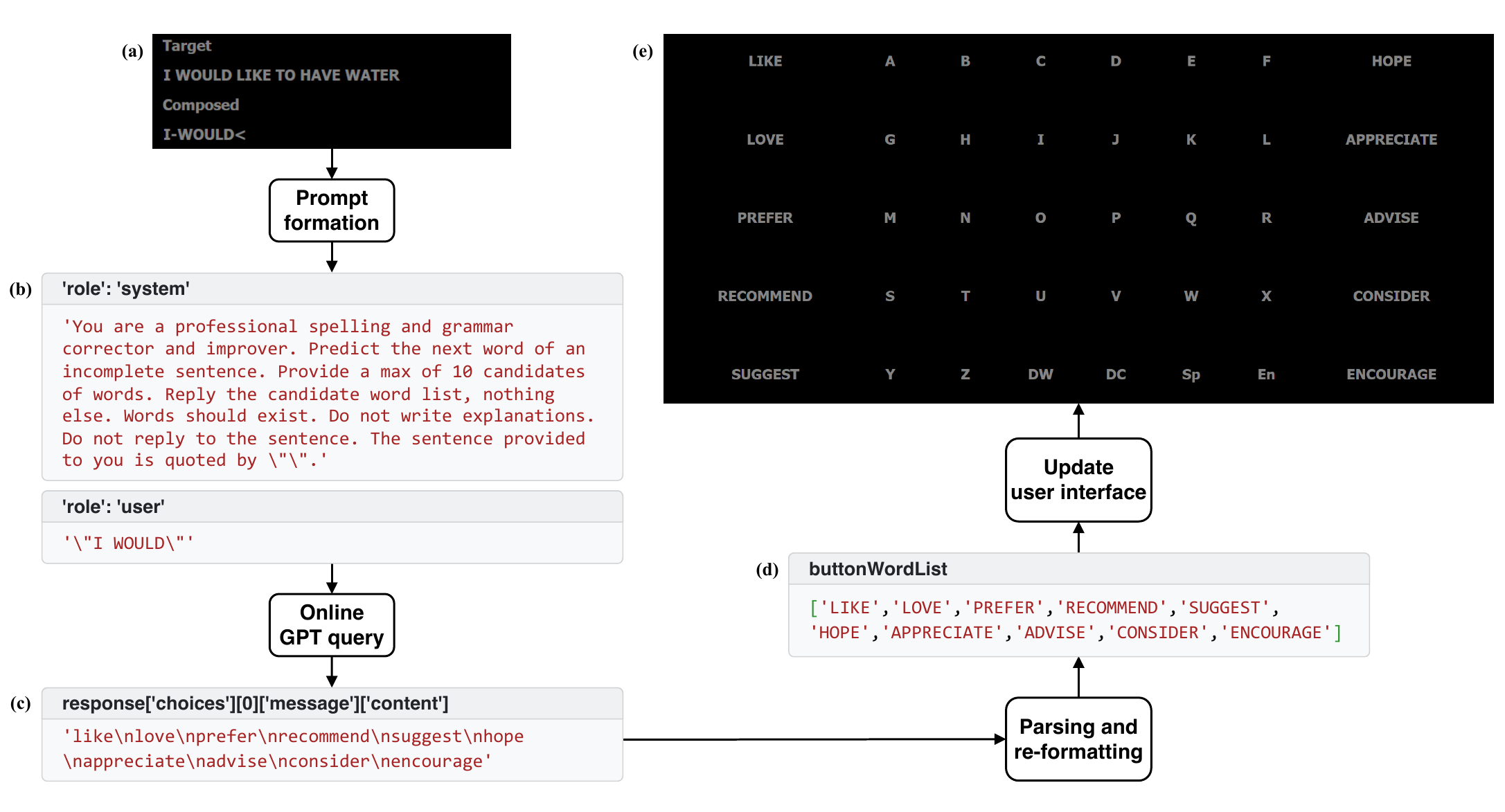} 
	\caption{\small Diagram of the GPT-3.5 query process in ChatBCI. \textbf{(a)}: The target sentence and the partially composed text by the user are shown. Spaces are displayed as ``-'' for visibility.  \textbf{(b)}: The list of two query messages taking the \textbf{system} and \textbf{user} roles, composed from the partial text. \textbf{(c)}: Response from the GPT-3.5-turbo API after sending the messages in \textbf{(b)}. \textbf{(d)}: List of candidate words formed by splitting and re-formatting the response string in \textbf{(c)}. \textbf{(e)}: GUI is updated with the candidate words.} 
	\label{fig:gptQuery}
\end{figure}

\subsection{Integration with Large Language Models (LLMs)}

\textbf{Predictive Spelling Through Remote GPT Query:} In ChatBCI, predictive spelling suggestions are generated from the partial text spelled by the user, through remote queries to the GPT-3.5-turbo API \cite{gpt3.5turbo}. Figure \ref{fig:gptQuery} illustrates an example of this process, where the user has composed ``I-WOULD'' as part of the target sentence of ``I-WOULD-LIKE-TO-HAVE-WATER'' (Figure \ref{fig:gptQuery}-(a)).

Given the partial text ``I-WOULD'', ChatBCI first generates a list of messages for remote GPT query. The messages are formed by filling the current partial text into a manually-engineered prompt template (Figure \ref{fig:gptQuery}-(b)). To ensure ChatBCI continuously provides helpful suggestions throughout the spelling process \cite{gptroles}, we utilize the role-taking functionality of the GPT-3.5-turbo API. The prompt is split into two messages, each assuming a distinct role. Specifically, the first message, in the \textbf{system} role, instructs GPT on the task, the number of predictive typing keys to include, and the required response format. The second message, in the \textbf{user} role, contains the partial text entered by the user.

After sending the query messages to GPT, ChatBCI receives a response string. Manual fine-tuning was employed to design the prompt template in Figure \ref{fig:gptQuery}-(b), so that most response strings from various partial text inputs contain valid typing suggestions that are separable using a consistent rule, as shown in Figure \ref{fig:gptQuery}-(c). The string is then split and reformatted into individual word suggestions (Figure \ref{fig:gptQuery}-(d)), and the BCI is updated with the word suggestions acquired from the current partial text input (Figure \ref{fig:gptQuery}-(e)). This allows the user to directly select ``LIKE'', thereby, significantly saving time and keystrokes compared to typing each letter individually. The entire process (Figure \ref{fig:gptQuery}) is repeated whenever a letter or word is added to or removed from the partial text, ensuring word suggestions remain updated based on the new partial text.\bigskip\\
\textbf{Word Completion and Word Prediction:} Using the prompt template shown in Figure \ref{fig:gptQuery}, GPT-3.5 intelligently provides two types of predictive spelling suggestions: \emph{word completion} when the last word is incomplete, and \emph{word prediction} when the last word is fully typed.  For example, in the text ``I-WANT-TO-B'' which ends with an incomplete word, GPT suggests predictions that complete the last word, such as ``BE'' or ``BUY''. On the other hand, for a text such as ``I-WOULD'' where the last word is complete but the sentence is not, GPT suggests predictions for the next word, for example ``LIKE'' or ``WANT''.

To integrate both scenarios of word completion and word prediction in ChatBCI, we designed $2$ simple rules for updating the composition based on word suggestions, optimizing intuitiveness and usability:
\begin{enumerate}
    \item We define the last word in a partially composed text as all characters after the last space character, if there is at least one space character in the text. Otherwise, the last word is the text. \medskip\\
    For example, for the text ``I'', the last word is ``I''; for the text ``I-WANT-TO-B'', the last word is ``B'';  for the text ``I-WOULD'', the last word is ``WOULD''; and the last word for the text ``I-WOULD-'' is empty.
 \\
    \item Whenever a candidate word on the GUI is selected, it always replaces the last word in the partially-composed text, and adds a space character to the end of the text.\medskip\\
  For example,  the result of selecting word ``BUY'' for the text ``I-WANT-TO-B'' is ``I-WANT-TO-BUY-'' (\emph{word completion}); and the result of selecting word ``LIKE'' for the text ``I-WOULD-'' is ``I-WOULD-LIKE-'' (\emph{word prediction}).
 \end{enumerate}
Each word selected from the suggestions automatically adds a space to the end of the text and triggers a new GPT query to update the suggestions based on the current text with the newly added word. This allows users to chain new words with just one key selection per word, thereby, significantly reducing keystrokes and speeding up the sentence composition process. Thus, the proposed ChatBCI can continuously provide flexible, dynamic predictive spelling suggestions, offering an effective means for composing long sentences.

%%%%%%%%%%%%%%%%%%%%%%%%%%%%%%
\subsection{P300 Detection and Target Character Recognition}
%%%%%%%%%%%%%%%%%%%%%%%%%%%%%%

To achieve P300-based spelling, the P300 component should be first detected from the acquired EEG signals, and then the corresponding target key needs to be identified. Determining the presence or absence of a P300 evoked potential from EEG signals can be treated as a binary classification problem, whereas target character recognition is a multi-class classification problem involving $40$ classes corresponding to the $5 \times 8$ grid keyboard (see Figure \ref{fig:3}). Here, we describe the algorithms used for these two tasks. \bigskip\\
\textbf{P300 Detection:} We considered the Stepwise Linear Discriminant Analysis (SWLDA) classifier for detecting the P300 evoked potential from the EEG signals. SWLDA has shown strong performance in prior P300 speller BCIs \cite{krusienski2008toward, speier2011natural, speier2015incorporating, mainsah2014utilizing, Chen2022Language, speier2018improving, chandravadia2022comparing, ryan2010predictive, kaufmann2012spelling, throckmorton2013bayesian,krusienski2006comparison}, making it a good candidate for ChatBCI.
 
SWLDA utilizes a stepwise feature selection process, iteratively adding or removing features based on their statistical significance and each feature's individual contribution to the model. This process involves ordinary least squares (OLS) regression for evaluating features. The feature selection begins with forward regression, where OLS estimates model parameters and assesses the significance of each feature. The feature with the lowest \textit{p}-value is added to the model first. In the forward selection phase, additional features demonstrating a significant and unique contribution to the variance (here with \textit{p}-values less than $0.1$) are included in the model. The subsequent backward elimination phase evaluates whether each included feature continues to contribute significantly, removing those with \textit{p}-values greater than $0.25$. This iterative process of forward selection and backward elimination continues until no further features meet the inclusion or exclusion criteria. The details of the Stepwise Feature Selection algorithm utilized in the SWLDA classifier are provided in the Appendix (Algorithm \ref{alg:1}). Features selected by the algorithm are used to calculate the Linear Discriminant Analysis (LDA) score to differentiate between the two classes, corresponding to the presence or absence of P300. We used the \texttt{LinearDiscriminantAnalysis} class implemented in scikit-learn version 1.3.1 \cite{scikit-learn}, with the default Singular Value Decomposition (SVD) solver. \bigskip\\
\textbf{Target Character Recognition:} As illustrated in Figure \ref{fig:3}, the keyboard panel of the GUI consists of $8$ columns and $5$ rows. We associated each column (left to right) with stimulus codes $1$ to $8$, and each row (top to bottom) to stimulus codes $9$ to $13$. The combination of column and row codes is used to identify the target character. For instance, [$6$, $11$] corresponds to the character ``Q''. Due to low SNR, we required $8$ repetitions to recognize the target character. Each repetition involved $13$ flashes, each corresponding to one of the $13$ stimulus codes.

Let \( s_i^{(r)} \) denote the LDA score for the stimulus code \( i \) ($i=1, \cdots, 13$) from the \( r \)-th repetition of total of \( n \) repetitions. \( S_i \) is defined as the cumulative sum of classifier scores for stimulus code \( i \) from the first to the \( n \)-th repetition. A cumulative score $S_i$ for each stimulus code across all repetitions ($n$) is obtained as
\begin{equation}
S_i = \sum_{r=1}^n s_i^{(r)}.
\end{equation}
The P300 response is triggered when the user focuses on a particular column or row flash. Therefore, the target character can be identified from the row and column with the highest cumulative scores as
\begin{equation}
\mbox{target column} = \mathop{\arg \max}_{i \in \{1:8\}} S_i, \qquad \mbox{target row} = \mathop{\arg \max}_{i \in \{9:13\}} S_i
\end{equation}

The intersection of the highest-scoring column and row gives the most likely target character, which is then selected.

%%%%%%%%%%%%%%%%%%%%%%%%%%
\subsection{Evaluation Metrics}\label{eval}
%%%%%%%%%%%%%%%%%%%%%%%%%%
The performance of the proposed ChatBCI was evaluated using a series of metrics. The accuracy was assessed using two measures of ``Selection Accuracy'' and ``Success Rate (SR)''. The speed was assessed using the measure of ``Time to Complete''. The performance was also evaluated using ``Information Transfer Rate'' which takes into account both speed and accuracy. The spelling efficiency was evaluated based on a few measures involving ``keystroke'' analysis. In what follows, we describe these metrics.\bigskip\\
%%%%%%%%%%%%%%%
\textbf{Time to Complete:} This metric represents the time that it takes to complete writing the intended sentence using the speller. A shorter time to complete writing, suggests a faster speller. \bigskip\\
%%%%%%%%%%%%%%
%%%%%%%%%%%%%%%
\textbf{Typing Speed:} This metric is defined as the ratio of total characters in the sentence over the time it takes to complete writing the sentence.
 \bigskip\\
%%%%%%%%%%%%%%
%%%%%%%%%%%%%%%
\textbf{Selection Accuracy:} Selection accuracy is calculated as the ratio of correct selections to the total number of selections used to type the sentence. High selection accuracy indicates that fewer attempts were needed for correct spelling. \bigskip\\
%%%%%%%%%%%%%%
%%%%%%%%%%%%%%%
\textbf{Success Rate (SR):} This metric is calculated as the ratio of correctly-typed characters to the total number of characters in the sentence. A high SR for a BCI, suggests its better capability to accurately capture the user's intent. This metric is calculated only for the copy-spelling task, as the target sentence is known. \bigskip\\
%%%%%%%%%%%%%%%
%%%%%%%%%%%%%%%
\textbf{Information Transfer Rate (ITR):} In letter-by-letter speller BCIs, the ITR is calculated in terms of bits/min following \cite{liu2018deep}
\begin{equation}
  \mbox{ITR} = \frac{B}{T/60},
  \label{itr}
\end{equation}
where
\begin{equation}\label{BB}
B=\left( (1 - P) \log_2 \frac{1-P}{N-1} + P \log_2 P + \log_2 N \right),
\end{equation}
and $N$ is the number of selections in the GUI, $P$ represents the probability of making a correct selection, and $T$ (in seconds) denotes the time required to make one selection. Note that in letter-by-letter spellers, each selection yields one character. 

However, in speller BCIs, such as the proposed ChatBCI, with predictive typing capability (e.g., word completion), a selection can produce multiple characters, therefore, the number of selections and characters may not be equal. To account for this difference, we calculate ITR\(^*\) for each completed sentence as 
\begin{equation}\label{eq:itr*}
\mbox{ITR}^\star = \frac{B}{T/60}\times \alpha,
\end{equation}
where $\alpha$ represents the average number of characters generated per selection defined as
\begin{equation}
\alpha=\frac{\# \mbox{of characters in the sentence}}{\# \mbox{of selections}}.
\end{equation}
For a letter-by-letter speller, $\alpha = 1$, making ITR\(^*\) in (\ref{eq:itr*}) equivalent to ITR in (\ref{itr}), and for a speller BCI with predictive spelling, where each selection typically outputs more than one character, $\alpha > 1$. 

Another difference between ChatBCI and other speller BCIs that impacts the calculation of information transfer rate, is the difference in the keyboard GUI.  A typical keyboard GUI consists of only characters and function keys, while the GUI in ChatBCI offers additional $10$ GPT-suggested words, with various number of characters (see Fig. \ref{fig:3}). This impacts $N$ in (\ref{BB}) for the calculation of information transfer rate. To address this issue, we consider two scenarios for calculating ITR\(^*\):
\begin{itemize}
 \item[-] $\mbox{ITR}^\star$-1:  We consider $N=28$, corresponding to the $26$ letters plus $2$ function keys (space and enter) available in the GUI (see Fig. \ref{fig:3}). This choice of $N$ provides the minimum number of selections available in the GUI, where no word suggestions by GPT are offered. Therefore,  {ITR}$^\star$-1 represents the lower bound for information transfer rate of ChatBCI.
 \medskip
 \item[-] $\mbox{ITR}^\star$-2: We consider $N=28+M$, corresponding to the sum of $26$ letters, $2$ function keys (space and enter), and the average number of characters in suggested words displayed in the GUI across all selections used to complete the sentence ($M$). {ITR}$^\star$-2 provides a good estimate of the observed information transfer rate. \end{itemize}
For both above scenarios, $T$ in (\ref{eq:itr*}) is calculated as follows. The duration of each flash is set to $40$ ms, followed by an inter-stimulus interval of $100$ ms. One sequence consists of $13$ flashes going over the $8$ columns and $5$ rows in the GUI, with each row and column flashing once in random order (thereby, ensuring the target character is flashed twice, once for the row and once for the column containing the character). There is a $1$-s  interval between sequences. After a key is selected, a $2$-s interval occurs. With $8$ repetitions required to make a selection, $T$ in (\ref{itr}) for ChatBCI is obtained as
\begin{equation}
\hspace{-10mm}T=2~\mbox{s} + (140 \mbox{ \small ms/flash} \times 13 \mbox{ \small flashes/seq.} + 1~\mbox{s/seq.}) \times 8 \mbox{ \small repetitions}=24.56~\mbox{s}. 
\end{equation}
\bigskip
%%%%%%%%%%%%%%%
%%%%%%%%%%%%%%%
\textbf{Keystroke Analysis:} To assess the predictive typing capability of ChatBCI, here, we propose to use keystroke analysis \cite{conijn2022early}. Using fewer keystrokes to complete a spelling task reflects a more efficient speller system, resulting in faster typing speeds and less user fatigue. This improvement can be measured quantitatively as keystroke savings (KS), defined as \cite{carlberger1997profet, newell1998role, garay2006text}
\begin{equation}
\mbox{\footnotesize{KS}}(\%) = \frac{\mbox{\footnotesize{Normal Keystrokes}} - \mbox{\footnotesize{Actual Keystrokes}}}{\mbox{\footnotesize{Normal Keystrokes}}} \times 100,
\label{ks}
\end{equation}
where ``Normal Keystrokes'' refers to the number of keystrokes needed to spell the sentence letter-by-letter, while ``Actual Keystrokes'' represents the keystrokes actually used by the speller system to complete the sentence. If corrective selections like deletions or re-selections are made, they are not included in the ``Actual Keystrokes'' count. This approach assesses the predictive efficiency of the speller, by assuming an ideal spelling process without typing mistakes or corrections \cite{trnka2008evaluating,trnka2009user}. Thus, only keystrokes that extend the longest common match between the spelled output and the target sentence are considered, ignoring any mistakes, corrections, or unnecessary inputs.

Speller systems may incorporate word prediction, completion, or a combination of both.  To better evaluate their KS performance, theoretical upper bounds based on ideal word completion (WC) or based on ideal word prediction (WP) have been suggested \cite{trnka2008evaluating} \footnote{We have excluded the one extra keystroke in these two theoretical bounds used in \cite{trnka2008evaluating} to only focus on the keystroke savings based on words achieved by the system.}. In a system with an ideal WC, the minimum required keystrokes for each word is two—one for typing the first letter and one for selecting the predicted word. Based on this definition, the theoretical maximum KS with ideal WC (KS-WC\textsubscript{max}) is obtained as 
\begin{equation}
\hspace*{-25mm}\mbox{\footnotesize{KS-WC}}_{\max}(\%) = \frac{\mbox{\footnotesize{Number of Characters in Sentence}} - 2\times\mbox{\footnotesize{Number of Words in Sentence}}}{\mbox{\footnotesize{Number of Characters in Sentence}}} \times 100,
\label{kswc}
\end{equation}
Similarly, in a system with an ideal WP, the minimum required keystrokes is one— one for selecting it.  Based on this definition, the theoretical maximum KS with ideal WP (KS-WP\textsubscript{max}) is defined as
\begin{equation}
\hspace*{-25mm}\mbox{\footnotesize{KS-WP}}_{\max}(\%) = \frac{\mbox{\footnotesize{Number of Characters in Sentence}} - \mbox{\footnotesize{Number of Words in Sentence}}}{\mbox{\footnotesize{Number of Characters in Sentence}}} \times 100.
\label{kswp}
\end{equation}
\bigskip
To better assess how far the speller is from achieving the maximum theoretical keystroke savings, we introduce the keystroke savings deficit ratio (KS-DR), defined as
\begin{equation}
\mbox{\footnotesize{KS-DR}}(\%) = (1-\frac{\mbox{\footnotesize{KS}}}{\mbox{\footnotesize{KS-WP}}_{\max}})\times 100.
\label{ksdr}
\end{equation}
A lower KS-DR indicates that the speller is closer to an ideal predictive system, making KS-DR a useful target for optimization and performance enhancement of the speller.

Here, we provide an example of how the KS analysis is applied for an example sentence ``I-WOULD-LIKE-TO-HAVE-WATER”. This sentence has $6$ words, and $26$ characters. For an ideal speller with word completion capability, the KS-WC\textsubscript{{max}} given by (\ref{kswc}) is $\frac{(26-2\times6)}{26}\times 100= 53.85\%$, while the KS-WP\textsubscript{{max}} given by (\ref{kswp}) is $\frac{(26-6)}{26}\times 100= 76.92\%$.

Now, assume a user spelled the same sentence using a speller BCI with predictive typing capabilities, through $10$ selections as 
 \begin{itemize}
 \footnotesize
    \item[-] Selection 1: ``I"
    \item[-] Selection 2: ``I-"
    \item[-] Selection 3: ``I-W"
    \item[-] Selection 4: ``I-WOULD-"
    \item[-] Selection 5: ``I-WOULD-LIKE-"
    \item[-] Selection 6: ``I-WOULD-LIKE-TO-"
    \item[-] Selection 7: ``I-WOULD-LIKE-TO-H"
    \item[-] Selection 8: ``I-WOULD-LIKE-TO-HAVE-"
    \item[-] Selection 9: ``I-WOULD-LIKE-TO-HAVE-W"
    \item[-] Selection 10: ``I-WOULD-LIKE-TO-HAVE-WATER"
\end{itemize}
With $10$ selections, using (\ref{ks}), KS is $\frac{(26-10)}{26}\times 100=61.54\%$, and KS-DR given by (\ref{ksdr}) is $(1-\frac{61.54}{76.92})\times 100=19.99\%$.

%%%%%%%%%%%%%%%%%%%%%%%%%%%%%%%%%%%%%
%%                                 %%
%% ------  Experiments  ------  %%
%%                                 %%
%%%%%%%%%%%%%%%%%%%%%%%%%%%%%%%%%%%%%

\section{Experiments}\label{sec:exp}

\subsection{Participants and Experimental Setup}
Seven healthy and right-handed volunteers (four males, three females, ages ranging from $18$ to $50$) were recruited for this study. All had normal or corrected-to-normal vision. Only one subject had prior experience with speller BCIs. Written informed consent forms, approved by the Rutgers Institutional Review Board (IRB), were obtained from each participant prior to the experiments.

\begin{figure}
    \centering 
    \includegraphics[width=0.55\textwidth]{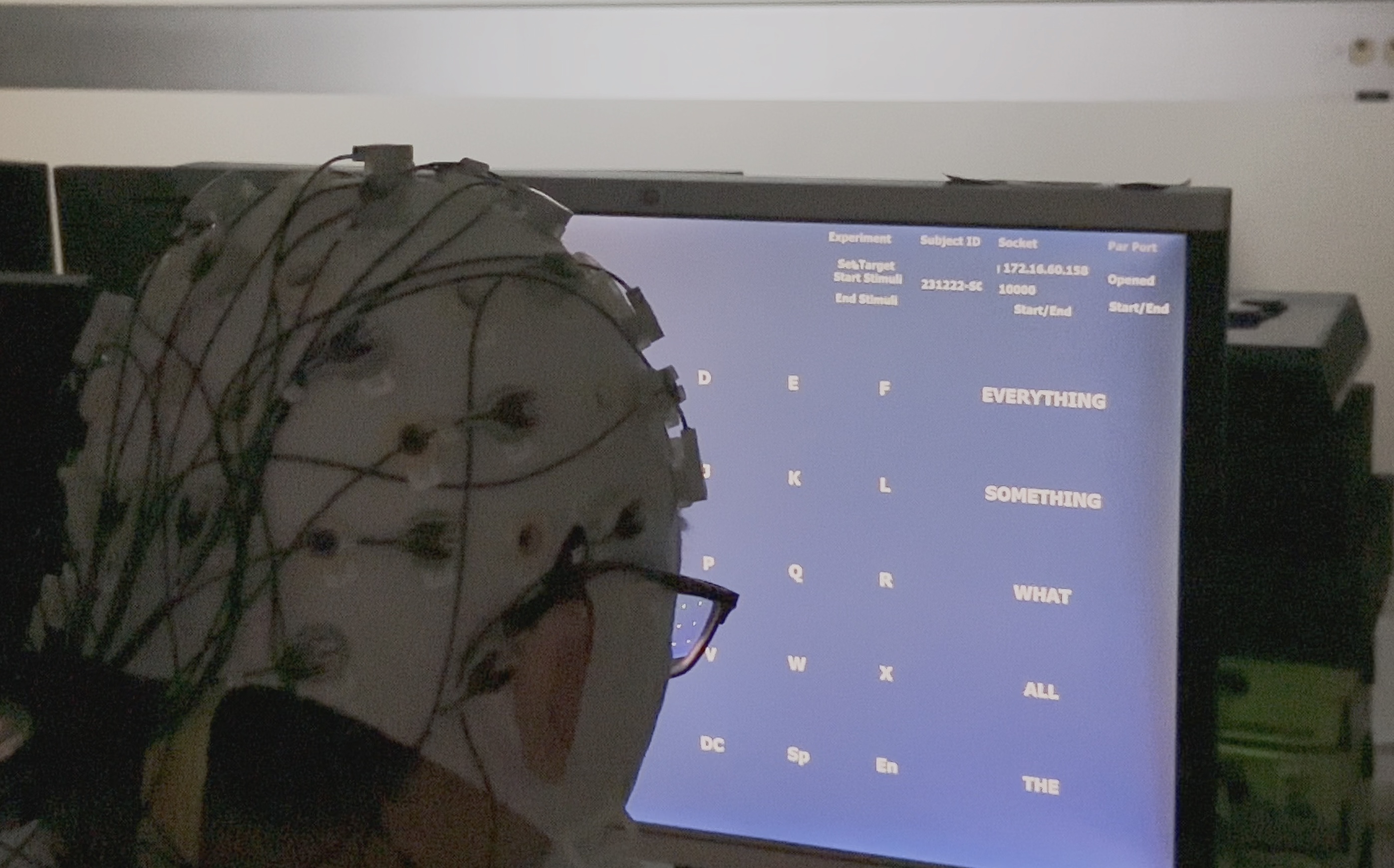} 
    \caption{\small Example of the experimental setup: a subject is using ChatBCI to type a sentence.} 
    \label{fig:env}
  \end{figure}

The experiment was conducted in a dimly-lit and quiet room (Figure \ref{fig:env}). Subjects were positioned approximately $60$ cm from the computer monitor. EEG signals were acquired using a $32$-channel EEG system (Brain Products) at a sampling rate of $250$ samples/s.  Participants were instructed to minimize movements during the experiment.

\begin{figure}
    \centering
    \begin{subfigure}[b]{0.23\linewidth}
        \includegraphics[width=\linewidth]{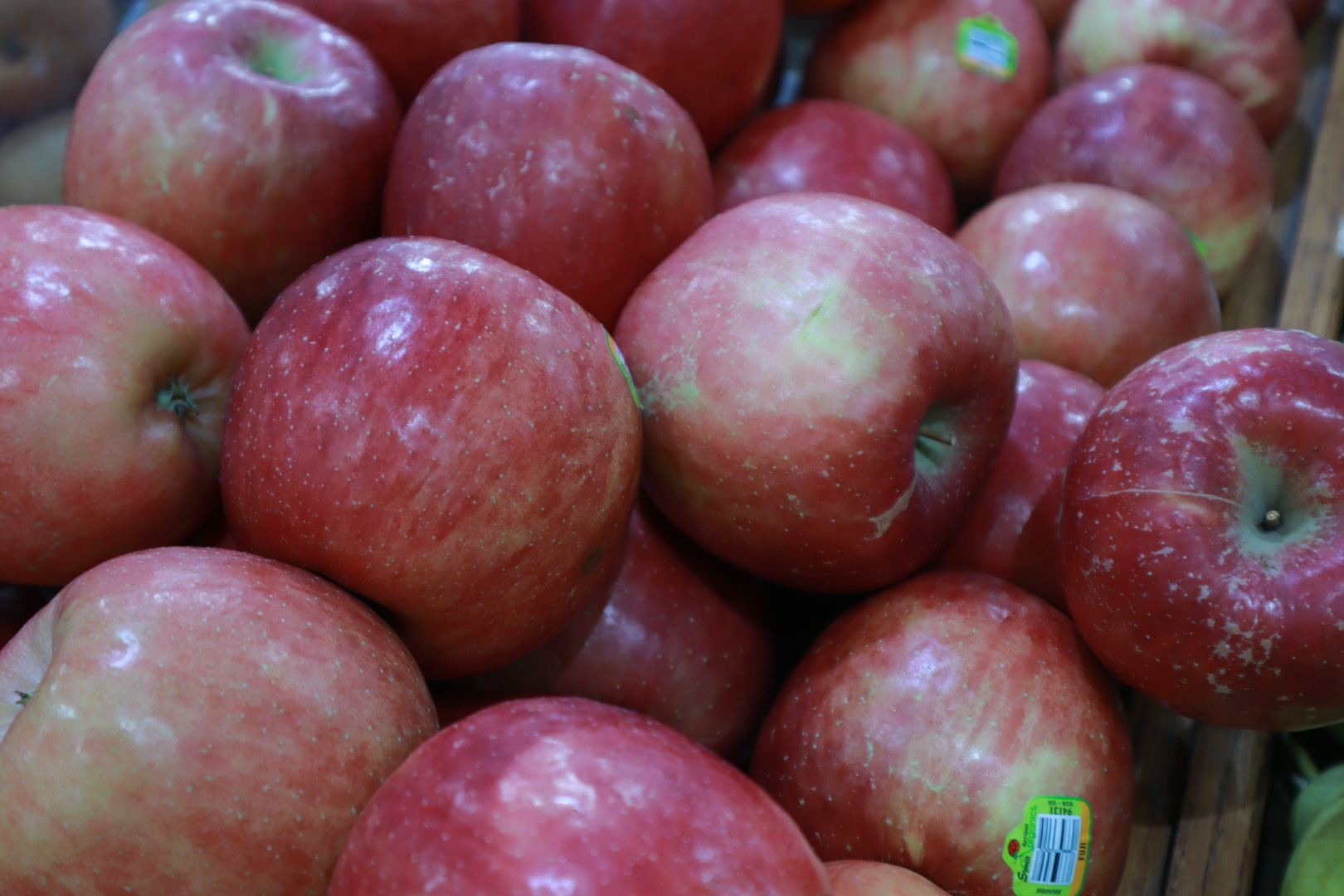}
        \caption{Apple}
        \label{fig5:a}
    \end{subfigure}
    \hfill 
    \begin{subfigure}[b]{0.23\linewidth}
        \includegraphics[width=\linewidth]{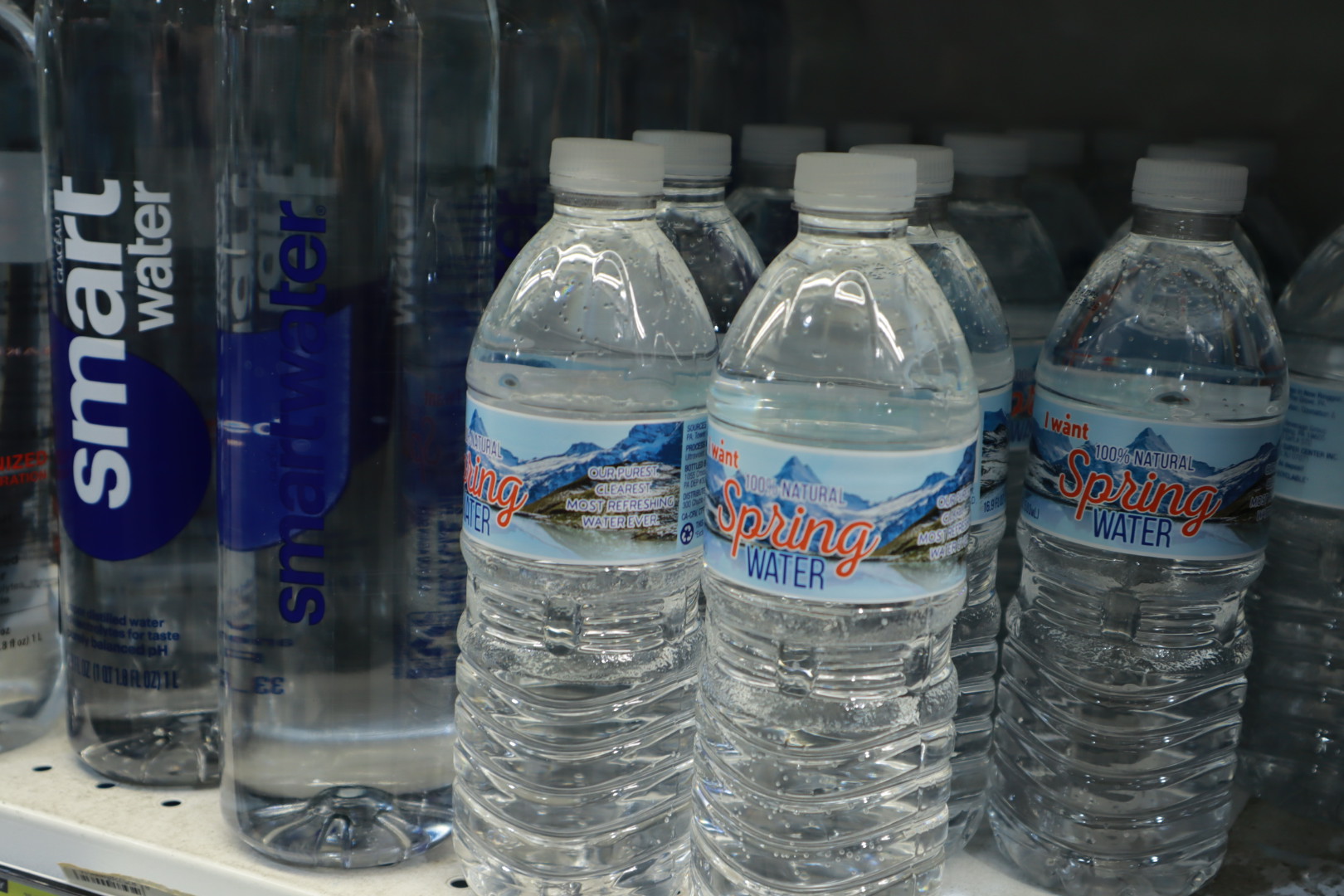}
        \caption{Water}
        \label{fig5:b}
    \end{subfigure}
    \hfill 
    \begin{subfigure}[b]{0.23\linewidth}
        \includegraphics[width=\linewidth]{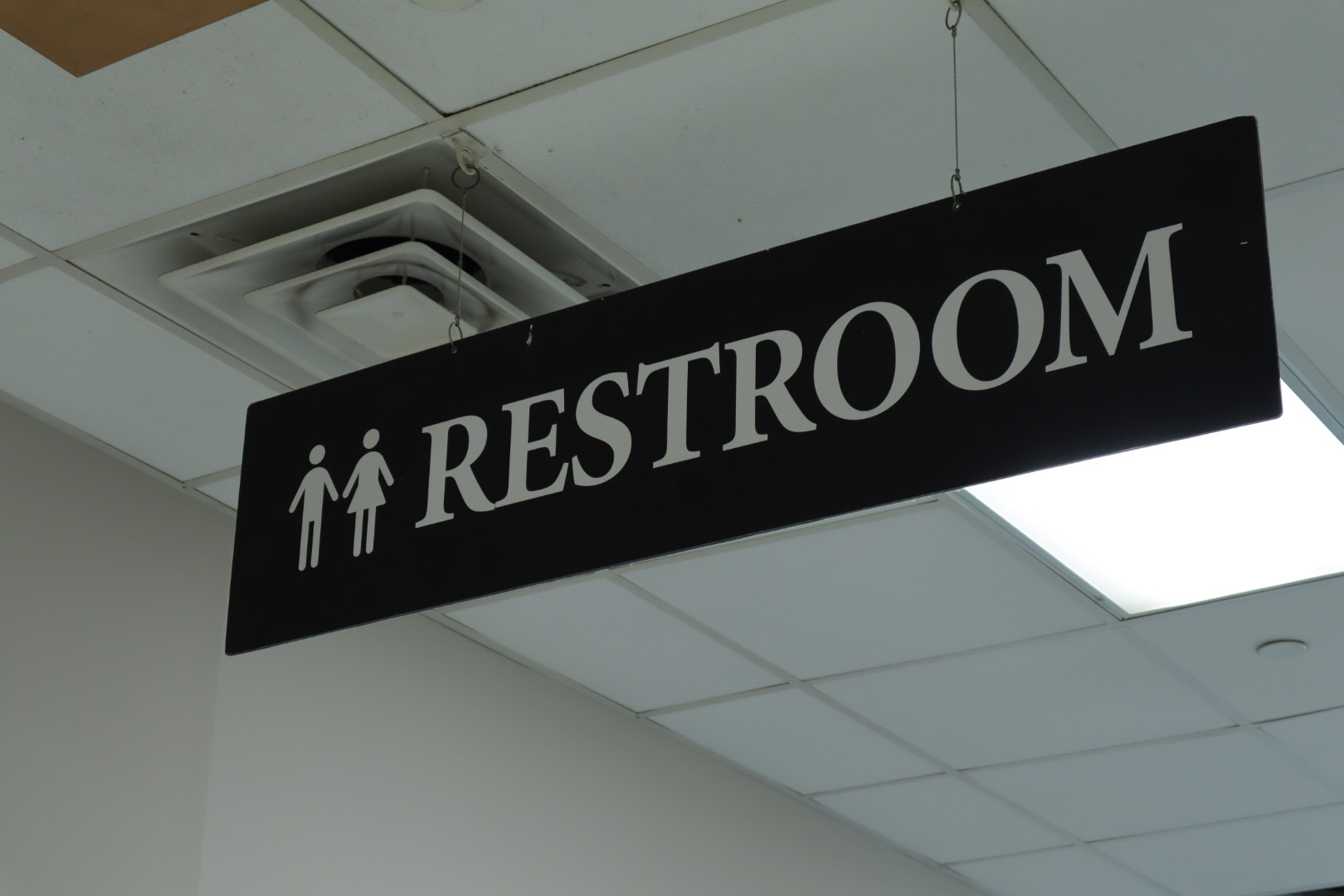}
        \caption{Restroom}
        \label{fig5:c}
    \end{subfigure}
    \hfill
    \begin{subfigure}[b]{0.23\linewidth}
        \includegraphics[width=\linewidth]{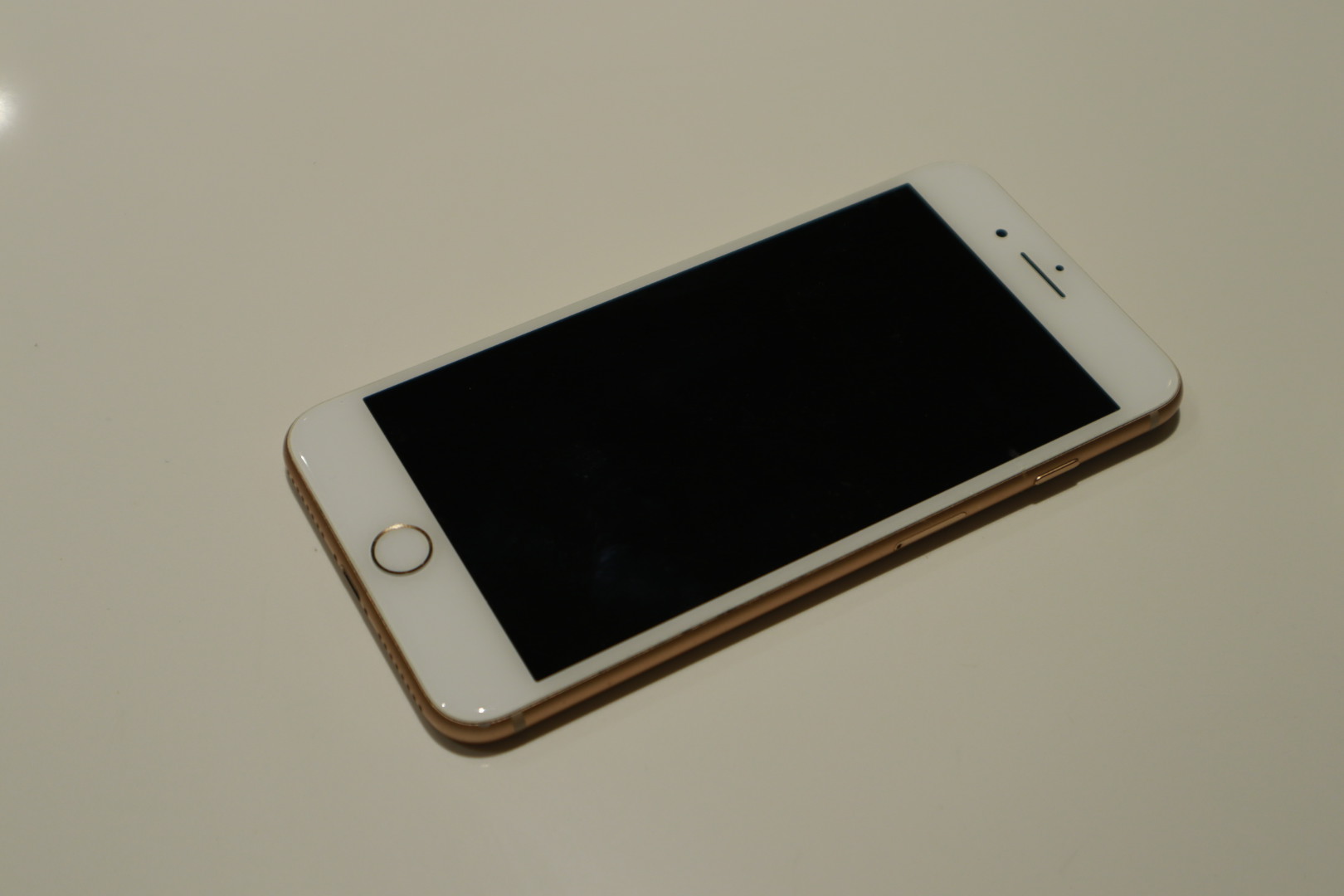}
        \caption{Phone}
        \label{fig5:d}
    \end{subfigure}

    \caption{\small Online session Task 1- Four images representing items relevant to daily life activities are presented to each subject. Subjects are asked to select one image and compose a meaningful sentence relevant to the selected image.}
    \label{fig:5}
\end{figure}

\subsection{Sessions}
Each subject participated in five sessions: a calibration (training) session, a validation session, and three online (testing) sessions. The online sessions included: 1) copy-spelling a self-created sentence using ChatBCI (Task 1), 2) a control task of copy-spelling the same sentence letter by letter, and 3) improvising a sentence with words suggested by ChatBCI (Task 2). Below we describe the details of these sessions.\bigskip\\
%%%%%% Calibration
\textbf{Calibration Session:} In this session, a total of $20$ keys were selected as calibration keys, ensuring a balanced distribution across the keyboard (Figure \ref{fig:3}). The selected keys were ``A'', ``En'', ``B'', ``Sp'', ``E'', ``Z'', ``F'', ``Y'', ``M'', ``R'', ``H'', ``W'', ``K'', ``T'', ``I'', ``V'', ``U'', ``J'', ``N'', ``Q''. 

Each subject was instructed to focus on the key highlighted in green, which remained green for $2$ s. Then, all rows and columns on the GUI flashed in yellow to evoke a P300 response. This process was repeated five times for each calibration key, and continued until all $20$ calibration keys were presented to the subject.  After the calibration session, a personalized SWLDA model was trained and used in all subsequent sessions to detect key selections for the subject. \bigskip\\
%%%%%% Validation
\textbf{Validation Session:} The validation session assesses the usability of the personalized model and helps the user become familiar with focusing on keys to make selections. A maximum of $10$ trials was allowed for each subject. During each trial, one of the first $10$ calibration keys lit up in green, indicating the key the subject should select. Then, the keyboard started flashing. After $8$ rounds of repeated flashes, the system determined the key selected by the user and checked whether it matches the target key for that trial. The subject must make $3$ consecutive correct selections within $10$ trials to pass the validation session; otherwise, the calibration and validation sessions must be repeated.  All seven participants in the study passed the validation session. \bigskip\\
%%%%%% Online Session: Task 1
\textbf{Online Session-Task 1:} Task 1 was a copy-spelling task, with the sentence created by the subjects. Each subject was presented with four images of ``Apple'', ``Water'', ``Restroom'', and ``Phone'', without any descriptions or labels (Figure \ref{fig:5}). Prior to the experiment, subjects were asked to select one image and compose a coherent sentence relevant to the selected image, excluding punctuation. This sentence was used as their target sentence, which was then entered into the ``Target'' block of the GUI, as reference (Figure \ref{fig:3}). Subjects were then asked to copy-spell the target sentence using ChatBCI, and were encouraged to employ the GPT-based word suggestions shown on the GUI during this process.  Subjects were instructed to correct mistakes, if they occurred, using as few key selections as possible by utilizing the ``DC'' (Delete Character) and ``DW'' (Delete Word) function keys. \bigskip\\
%%%%%% Online Session: Task 1-Control
\textbf{Online Session-Task 1 (Control):} To establish a control case, after completing the copy-spelling task, subjects were asked to write the same target sentence, \emph{letter by letter}, using only the alphabet and function keys. In this scenario, although the same ChatBCI GUI was used, only the alphabet and the function keys remained visible. The GPT query functionality was turned off, and the $10$ candidate word keys were hidden. \bigskip\\
%%%%%% Online Session: Task 2
\textbf{Online Session-Task 2:} Task 2 involved improvising a sentence using ChatBCI and represents a more realistic scenario for communication. Subjects were instructed to create a meaningful sentence starting with the letter ``H''. They were encouraged to avoid typing words letter by letter, and instead, utilize as many suggested words from ChatBCI as possible. While subjects were allowed to complete sentences without corrections, they were asked afterward to report the number of incorrect selections to assess selection accuracy.
%%%%%%%%%%%%%%%%%%%%%%%
%%% signal processing
%%%%%%%%%%%%%%%%%%%%%%%
\subsection{EEG Signal Processing}
The acquired EEG signals were filtered using a bandpass filter ($0.5$-$30$ Hz) to minimize noise and artifacts. Signals from $16$ channels (Fz, Cz, Pz, Oz, P3, P4, P7, P8, FC1, FC2, CP1, CP2, C3, C4, O1, O2) were selected for analysis. Data from each channel in the interval of $0$ ms to $700$ ms post-stimulus, following each flash, was extracted, resulting in a feature matrix with the dimension of $175 \times 16$ per flash. To reduce feature dimension, the moving average technique with an average window and decimation factor of $12$ for each channel was applied, resulting in a feature vector length of $240$ ($175/12$, using the ceiling function $\times 16$) per flash. These feature vectors were then used as input to the stepwise feature selection for the SWLDA classifier.
%%%%%%%%%%%%%%%%%%%%%%%%%%%%%%%%%%%%%%%%%%%%%%
%%                                          %%
%% ------  Results ------  %%
%%                                          %%
%%%%%%%%%%%%%%%%%%%%%%%%%%%%%%%%%%%%%%%%%%%%%%
\section{Results}\label{results}
In this section, we present the performance results for ChatBCI based on data obtained from the online tasks using the evaluation metrics described in Section \ref{eval}.

%%%%%% Online Session: Task 1
\subsection{Performance Analysis of ChatBCI-Task 1}
In Task 1, subjects were instructed to create a sentence using a selected image, and copy-spell the sentence using ChatBCI and also letter-by-letter. 
\subsubsection{\underline{Speed and Accuracy:}}

Table \ref{tab:2} summarizes the sentences each subject created from their selected image and used in the copy-spelling task (Task 1) and its control letter-by-letter version. The number of characters (\# of Char.) in each sentence, the number of selections required to complete writing each sentence, the time to complete the task, typing speed,  selection accuracy, and success rate (SR) are also summarized for each subject. 

\begin{table} 
    \centering
    \caption{\small Performance results of ChatBCI from Task 1 and its control version (letter-by-letter (LL)), for each subject.
 Average results across subjects are also reported, with bold numbers indicating the better result between ChatBCI and LL control case.}
    \label{tab:2}
    \resizebox{\textwidth}{!}{
\begin{threeparttable}
    \begin{tabular}{c | c | p{9.3cm}| c | c c | c c | c c | c c }
        \hline
        \textbf{} & \textbf{} & \textbf{} & \textbf{\#} & \multicolumn{2}{c|}{{\textbf{\# of}}} & \multicolumn{2}{c|}{\textbf{Time to Complete (min)}}
  & \multicolumn{2}{c|}{\textbf{Selection}} & \multicolumn{2}{c}{\textbf{SR}}\\
        \textbf{Subject}          & \textbf{Image}     & \textbf{Sentence}       & \textbf{of} & \multicolumn{2}{c|}{\textbf{{Selections}}} & \multicolumn{2}{c|}{\textbf{(Typing Speed (characters/min))}}  & \multicolumn{2}{c|}{\textbf{Accuracy (\%)}} & \multicolumn{2}{c}{\textbf{(\%)}}  \\
        \cline{5-12}
        \textbf{}          & \textbf{}     & \textbf{}     & \textbf{Char.} & \textbf{ChatBCI} &  \textbf{LL}& \textbf{ChatBCI} &  \textbf{LL}&  \textbf{ChatBCI} &  \textbf{LL} & \textbf{ChatBCI} &  \textbf{LL}   \\
        \hline

        S01 & Phone & I-WANT-TO-BUY-A-NEW-PHONE & 25 & 22 & 46 & 13.62 (1.84) & 28.53 (0.88) & 77.27 & 76.09 & 100 &100 \\
        \hline
        S02 & Phone & I-WOULD-LIKE-TO-CALL-MY-MOM & 27 & 17 & 41 &10.50 (2.57) &25.42 (1.06) &{94.12} &{82.93} & 100 &100 \\
        \hline
        S03 & Water & I-WANT-SOME-WATER & 17  & 25 & 23 &16.10 (1.06) &14.25 (1.19) & 72.00 &86.96 & 100& 100  \\
        \hline
        S04 & Water & I-JUST-HAD-WATER & 16 & 14 & 32 &8.65 (1.85)  &22.32 (0.72) & 85.71 &68.75 &100 & 87.50  \\
        \hline
        S05 & Restroom & I-WANT-TO-GO-TO-THE-RESTROOM & 28   & 13 & 65 &8.05 (3.48)&40.33 (0.69) &92.31 & {61.54} &100 &75.00  \\
        \hline
        S06 & Apple & AN-APPLE-A-DAY-KEEPS-DOCTORS-AWAY & 33 & 14 & 39 &8.70 (3.79) &24.18 (1.36) &100 &92.31 &100 &100  \\
        \hline
        S07 & Apple & THERE-ARE-SOME-APPLES-IN-THE-MARKET & 35 & 12 & 60 &7.40 (4.73) &37.85 (0.92) &100 &76.67 & 100 &100 \\
        \hline 
        \multicolumn{3}{c|}{\textbf{Average}} & {26\tnote{a}} & {\textbf{17\tnote{a}}} &{44\tnote{a}} & {\textbf{10.43}} (\textbf{2.76}) & {27.55} (0.97) & {\textbf{88.77}} & {77.89} & \textbf{100} & 94.64 \\
        \hline
    \end{tabular}%
      \end{threeparttable}
   }
\raggedright \tiny $^{a}$The number was rounded up using the ceiling function.
\end{table}

\begin{table}\small 
    \centering
    \caption{\small Information transfer rate for ChatBCI from Task 1 for each subject. For ChatBCI, two scenarios are considered ITR$^{*}$-1 (N=28), and  ITR$^{*}$-2 (N=28+M), where $M$=the average number of characters in suggested words displayed in the GUI across all selections used to complete the sentence.}
    \label{tab:itrtask1}
    \resizebox{0.75\textwidth}{!}{
\begin{threeparttable}
    \begin{tabular}{ c| c| c |c  }
        \hline
        \textbf{} & \textbf{LL}   & \multicolumn{2}{c} {\textbf{ChatBCI}}\\
       \cline{2-4}
        \textbf{Subject} & ITR$^{*}$ (bits/min)&  ITR$^{*}$-1 (bits/min)& ITR$^{*}$-2 (bits/min), ($M$)\\
        \textbf{} & $N=28$ & $N=28$ & $N=28+M$\\
        \hline
        S01 & 6.38 &13.35 & 16.99, ~(41.52)\\
        \hline
        S02 &  7.73 &18.65  &  N/A   \\
        \hline
        S03 &   8.68 & 7.99 & 10.52, ~(52.75) \\
        \hline
        S04 &  4.48 & 13.42 &  16.76, ~(36.15) \\
        \hline
        S05 &   2.95 & 25.3 &  32.54, ~ (44.75) \\
        \hline
        S06 &   6.85 & 34.25 & 35.31, ~ (42.15) \\
        \hline
        S07 &   9.94 & 27.68 &  45.57, ~ (56.18) \\
        \hline 
     {\textbf{Average}} & 6.72 & 20.09 & 26.28, ~(45.58) \\
        \hline
    \end{tabular}%
      \end{threeparttable}
   }
\end{table}

Results in Table \ref{tab:2} show that ChatBCI, with GPT integration, offers a significantly improved typing speed compared to the letter-by-letter control. On average, subjects took $10.43$ minutes to complete a sentence with ChatBCI, compared to $27.55$ minutes using the letter-by-letter approach—a $62.14\%$ reduction in time. ChatBCI also offered an average improvement of $184.54\%$ characters/min in typing speed compared to the letter-by-letter case. All subjects, except S03 (who had low selection accuracy), completed the task faster with ChatBCI compared to using letter-by-letter approach.

The accuracy is also improved as a result of incorporating GPT, as the average selection accuracy with ChatBCI is $88.77\%$, which is higher than the $77.89\%$ achieved using the letter-by-letter approach. This suggests that the predictive capabilities of ChatBCI contribute to more accurate selections. The SR is consistently $100$\% for all subjects when using ChatBCI, compared to an average of $94.64\%$ for the letter-by-letter approach. In fact, using the letter-by-letter approach, among the seven subjects, S04 was unable to complete the sentence due to two uncorrected characters, and S05 had an incomplete sentence because of reduced concentration from prolonged attempts (hence their SR is less than $100\%$). 

\subsubsection{\underline{Information Transfer Rate:}}
Table \ref{tab:itrtask1} summarizes the information transfer rate results for the ChatBCI and its control letter-by-letter (LL) version.  We considered \textit{P} in (\ref{BB}) to be equal to SR. As discussed earlier, two scenarios are considered for the calculation of ITR$^{*}$ for ChatBCI considering its predictive capabilities and its new GUI:  ITR$^{*}$-1 in which $N=28$, and and ITR$^{*}$-2 in which $N=28+M$, where $M$ represents the average number of characters in suggested words displayed in the GUI across all selections, used to complete the sentence. 

As can be seen, the information transfer rate is substantially higher with ChatBCI, averaging $20.09$ bits/min (for ITR$^{*}$-1, the lower bound), compared to $6.72$ bits/min with the letter-by-letter version (an improvement of $198.96\%$). If the GUI design is taken into account, this improvement goes up to $291.07\%$ for ITR$^{*}$-2. ITR$^{*}$ varies across subjects due to individual variability and the need for error correction, which may result in making several selections to output one character. Note that the log file of suggested words were not correctly recorded for S02, therefore, $M$ is not available for this subject. The highest ITR$^{*}$ with ChatBCI was achieved by subjects S06 and S07, due to their $100\%$ selection accuracy (see Table \ref{tab:2}), which resulted in minimizing the correction time involving character deletion and re-selection. 

\begin{table*}
    \centering
    \caption{\small Keystroke analysis for Online Session-Task 1. Words and letters in each sentence are color-coded, depending on whether they were written using word completion (WC), word prediction (WP), or were typed letter by letter (LL). WC: gray, WP: blue, LL: black}
    \label{tab:3}
    \resizebox{\textwidth}{!}{
    \begin{threeparttable}
\begin{tabular}{c|p{9.5cm}||c||c|c|c|c|| c|c|c ||c }
    \hline
    \textbf{Subject} & \textbf{Sentence} & \textbf{\# of} &\textbf{\# of} & \textbf{\# of} & \textbf{\# of} & \textbf{\# of} & \textbf{KS} & \textbf{KS-WC$_{\max}$} & \textbf{KS-WP$_{\max}$} & \textbf{KS-DR}  \\
    \textbf{} &  \textbf{}&\textbf{Keystrokes} & \textbf{Words} & \textbf{WCs} & \textbf{WPs} &  \textbf{LLs} &\textbf{(\%)} & \textbf{(\%)} & \textbf{(\%)} & \textbf{(\%)}  \\   
    \hline
        S01 & I-W\WC{ANT-}\WP{TO-}B\WC{UY-}\WP{A-}N\WC{EW-}\WP{PHONE} & 11& 7 & 3 & 3 & 1 & 56.00 & 44.00  & 72.00  & 22.22 \\ \hline
        S02 & I-WOU\WC{LD-}\WP{LIKE-}\WP{TO-}C\WC{ALL-}MY-M\WC{OM}& 15& 7 & 3 & 2 & 2& 44.44  & 48.15 & 74.07  & 40.00 \\ \hline
        S03 & I-W\WC{ANT-}SOME-W\WC{ATER} & 11 & 4 & 2 & 0 & 2& 35.29 & 52.94  & 76.47  & 53.85  \\ \hline
        S04 & I-JU\WC{ST-}H\WC{AD-}W\WC{ATER} & 9 & 4 & 3 & 0 &  1& 43.75  & 50.00  & 75.00  & 41.67 \\ \hline
        S05 & I-\WP{WANT-TO-GO-}TO-\WP{THE-}R\WC{ESTROOM} & 11 &  7 & 1 & 4 & 2 & 60.71  & 50.00  & 75.00  & 19.05 \\ \hline
        S06 & AN-\WP{APPLE-}A-\WP{DAY-KEEPS-}D\WC{OCTORS-}\WP{AWAY} &11 &  7 & 1 & 4 & 2 & 66.67 & 57.58  & 78.79  & 15.38 \\ \hline
        S07 & T\WC{HERE-}\WP{ARE-SOME-}AP\WC{PLES-}IN-\WP{THE-MARKET} &12 & 7 & 2 & 4 & 1& 65.71  & 60.00 & 80.00  & 17.86 \\ 
    \hline
    \multicolumn{2}{c|}{\textbf{Average}}& 12\tnote{a} & 7\tnote{a} & 3\tnote{a} & 3\tnote{a} &  2\tnote{a} & 53.22 & 51.81  & 75.90  & 30.00\\
    \hline
\end{tabular}
\end{threeparttable}
   }
\raggedright \tiny $^{a}$Numbers were rounded up using the ceiling function.
\end{table*}

\subsubsection{\underline{Keystroke Savings Analysis:}}
Table \ref{tab:3} summarizes the results of the keystroke analysis for ChatBCI in Task 1. Results are provided for each subject, along with averages across all subjects. For each subject's sentence, the table reports the number of keystrokes, the numbers of occurrences of word completion (WCs) and word prediction (WPs) events, the number of words written letter-by-letter (LL), keystroke savings (KS), theoretical KS-WC$_{\max}$ and KS-WP$_{\max}$, and KS-DR. Averages across subjects are also reported.

As can be seen, most subjects utilized a combination of word completion and word prediction techniques in their typing process. Subjects S01, S05, S06, and S07 demonstrated excellent performance, taking advantage of word prediction and completion capabilities of ChatBCI, achieving keystroke savings that exceeded theoretical KS-WC$_{\max}$, with keystroke savings of $56.00\%$, $60.71\%$, $66.67\%$, and $65.71\%$, respectively. Generally, increased use of word prediction features led to higher keystroke savings and lower KS-DR, as expected, bringing the speller closer to its theoretical KS-WP$_{\max}$.  The lowest KS-DR is achieved by Subject S06, with an impressively low KS-DR of 15.38\%.

Comparatively, Subjects S03 and S04 showed the worst performance in terms of keystroke savings, which can be attributed to their lack of use of word prediction features in their writing process ($0$ occurrences of word prediction events). The next worst performance belongs to Subject S02. While this subject had used word prediction for two words in the sentence, three words were completely or partially spelled letter-by-letter, increasing the overall number of keystrokes.

Overall, using word prediction features and minimizing letter-by-letter inputs improve keystroke savings and bring the system closer to its maximum performance. The proposed ChatBCI, by offering word completion, word prediction, and letter-by-letter typing, uniquely adapts to individual preferences, needs, and typing styles, with the potential to significantly reduce keystrokes.

%%%%%%%%%%%%%%%%%%%%%%%%%%%
%%%%%%%%%%%%%%%%%%%%%%%%%%%
\subsection{Performance Analysis of ChatBCI-Task 2}

In Task 2, subjects were instructed to improvise a meaningful sentence starting with the letter ``H'', utilizing the words that appear in the GUI. 

\subsubsection{\underline{Speed and Accuracy:}}
The improvised sentences composed by each subject in the online Task 2, along with the number of characters (\# of Char.) in each sentence, the number of selections, the time required to complete the task, and selection accuracy are reported in Table \ref{tab:4}.

\begin{table*}
    \centering
    \caption{\small Performance results of ChatBCI from Task 2, for each subject. Average results across subjects are also reported.}
    \label{tab:4}
    \resizebox{\textwidth}{!}{
    \begin{threeparttable}
\begin{tabular}{c|p{10cm}|c|c|c|c|cc}
    \hline
    \textbf{Subject} & \textbf{Sentence} & \textbf{\# of} & \textbf{\# of} &\textbf{Time to Complete (min)} & \textbf{Selection} & \textbf{ITR$^*$}-1 (bits/min)&  \textbf{ITR$^*$}-2 (bits/min), (M) \\
    \textbf{} &  \textbf{}&\textbf{Char.}&\textbf{Selections} & \textbf{(Typing Speed (characters/min))} & \textbf{Accuracy (\%)} & ($N=28$) & ($N=28+M)$ \\  
    \hline
        S01 & HIS-FRIENDS-WERE-CARING-SUPPORTIVE-AND-LOYAL & 44 & 7 & 4.28 (10.28) & 85.71 & 73.82&  94.44, (43) \\ \hline
        
        S02 & HERE-IT-BEGINS-WHERE-THEY-COMMENCE-TO-UNDERTAKE-THEIR-ADVENTURE & 63 & 11 & 8.02 (7.86) & 81.82 & 67.26 & 85.38,(40.7)\\ \hline
        
        S03 & HE-HAS-GONE-TOO-FAR-AWAY-NOW-AND-HE-HAS-RETURNED & 48 &13 & 9.25 (5.19)& 92.31 & 43.36 &  53.99, (35.33) \\ \hline
        
        S04 & HOPE-IS-NEVER-LOST  & 18 & 5 &  3.05 (5.90)& 100 & 42.28 & 52.87, (36.50) \\ \hline
        
        S05 &  HAD-AN-AMAZING-CONVERSATION-LAST-NIGHT-WITH-HIM-ABOUT-LIFE-AND-THE-FUTURE-UNCERTAINTIES & 87 & 13 & 8.63 (10.08) & 92.31 & 78.60 & 103.51, (52.50) \\ \hline
        
        S06 & HOME-DECOR-MAGAZINE-SUBSCRIPTION-TRENDS-DESIGN-IDEAS-FROM-MODERN-RENOVATIONS-AND-INTERIOR-DESIGN-LAYOUTS & 104 & 15 &9.25 (11.24) & 100 & 81.43 &  107.49, (53.36) \\ \hline
        S07 & HAS-NOT-FINISHED-YET-BUT-THEY-WILL-ACCOMPLISH-THE-MISSION-EVENTUALLY & 68 & 12 & 7.40 (9.19) & 100 & 66.55 & 86.57, (48.27)  \\
    \hline
    \multicolumn{2}{c|}{\textbf{Average}} & 62\tnote{a} & 11\tnote{a} & 7.13 (8.53) & 93.16 & 64.76 &  83.46, (44.24)\\
    \hline
\end{tabular}
\end{threeparttable}
   }
\raggedright \tiny $^{a}$Numbers were rounded up using the ceiling function.
\end{table*}

The number of characters in the improvised composed sentences ranged from $18$ to $104$, with an average of $62$ characters. The time to complete the task varied from $3.05$ minutes to $9.25$ minutes across subjects, averaging $7.13$ minutes. The maximum achieved typing speed was $11.24$ characters/min. Comparing the number of selections used to write each sentence with the number of characters that exist in each sentence, it is evident that significantly fewer attempts were required to complete writing the sentences. This is due to WP and WC features of the ChatBCI, allowing each selection to output multiple characters. On average, with only $11$ selections, $62$ characters were typed, resulting in an average character per selection of $5.64$, representing a $464$\% improvement compared to an ideal letter-by-letter speller without correction. Selection accuracy was also generally high among subjects, with an average of $93.16\%$. 

Compared to the online copy-spelling task (Task 1, Table \ref{tab:2}), while the average number of characters is much larger in Task 2 ($26$ vs $62$), ChatBCI completes Task 2 faster (average of $10.43$ minutes vs $7.13$ minutes) with a higher selection accuracy ($88.77$\% vs $93.16$\%).  This suggests that in realistic scenarios when users compose sentences in real time, the WC and WP features of the ChatBCI can significantly improve the performance of BCI speller.

\subsubsection{\underline{Information Transfer Rate:}}
Table \ref{tab:4} also summarizes the information transfer rate results for the ChatBCI from Task 2. For this task, since subjects will always make a selection based on the presented words we assumed the \textit{P} in (\ref{BB}) to be 1. When considering $N=28$ (ITR$^{*}$-1, the lower bound), ChatBCI achieves an impressive average information transfer rate of $64.76$ bits/min. If the number of characters available in the suggested words in the GUI are also considered (i.e., $N=28+M$, ITR$^{*}$-2), ChatBCI achieves an average information transfer rate of $83.46$ bits/min. Compared to copy-spelling task (Task 1) and its control letter-by-letter case (Table \ref{tab:itrtask1}), ChatBCI achieves a significantly improved information transfer rate in the improvised task (Task 2).

\begin{table*}
    %\centering
    \caption{\small Keystroke analysis for Online Session-Task 2 for writing improvised sentences with ChatBCI.  Words and letters in each sentence are color-coded, depending on whether they were written using word completion (WC), word prediction (WP), or were typed letter by letter (LL). WC: gray, WP: blue, LL: black, Multi-word prediction: red.}
    \label{tab:5}
    \begin{threeparttable}
    \resizebox{\textwidth}{!}{
\begin{tabular}{c|p{9cm}||c||c|c|c|| c|c|c ||c }
    \hline
    \textbf{Subject} & \textbf{Sentence} & \textbf{\# of} &\textbf{\# of} & \textbf{\# of} & \textbf{\# of} & \textbf{KS} & \textbf{KS-WC$_{\max}$} & \textbf{KS-WP$_{\max}$} & \textbf{KS-DR}  \\
    \textbf{} &  \textbf{}&\textbf{Keystrokes} & \textbf{Words} & \textbf{WCs} & \textbf{WPs} &  \textbf{(\%)} & \textbf{(\%)} & \textbf{(\%)} & \textbf{(\%)}  \\   
    \hline
        S01 & H\WC{IS-}\WP{FRIENDS-WERE-CARING-SUPPORTIVE-}\MWP{AND-LOYAL} & 7 & 7  & 1 & 5\tnote{+}  & 84.09 & 68.18 & 84.09 & 0 \\ \hline
        
        S02 & H\WC{ERE-}\WP{IT-BEGINS-WHERE-THEY-COMMENCE-TO-UNDERTAKE-THEIR-ADVENTURE} & 11 & 10 & 1 & 9   & 82.54 & 68.25 & 84.13 & 1.89  \\ \hline
        
        S03 & H\WC{E-}\WP{HAS-}G\WC{ONE-}\WP{TOO-FAR-AWAY-NOW-AND-HE-HAS-RETURNED} & 13 & 11 & 2 & 9   & 72.92 & 54.17 & 77.08 & 5.40 \\ \hline
        
        S04 & H\WC{OPE-}\WP{IS-NEVER-LOST}  & 5 & 4  & 1 & 3   & 72.22 & 55.56 & 77.78 & 7.15 \\ \hline
        
        S05 &  H\WC{AD-}\WP{AN-AMAZING-CONVERSATION-LAST-NIGHT-WITH-HIM-ABOUT-LIFE-}\MWP{AND-THE-FUTURE-}\WP{UNCERTAINTIES} &  13 & 14 & 1 & 11\tnote{+} & 85.06 & 67.82 & 83.91 & -1.37 \\ \hline
        
        S06 & H\WC{OME-}\WP{DECOR-MAGAZINE-SUBSCRIPTION-TRENDS-DESIGN-IDEAS-FROM-MODERN-RENOVATIONS-AND-INTERIOR-DESIGN-LAYOUTS} & 15 & 14 & 1 & 13  & 85.58 & 73.08 & 86.54 & 1.11 \\ \hline
        S07 & H\WC{AS-}\WP{NOT-FINISHED-YET-BUT-THEY-WILL-ACCOMPLISH-THE-MISSION-EVENTUALLY} & 12 & 11 & 1 & 10  & 82.35 & 67.65 & 83.82 & 1.75 \\
    \hline
    \multicolumn{2}{c|}{\textbf{Average}}& 11\tnote{a} & 11\tnote{a} & 2\tnote{a}  & 9\tnote{a} & 80.68 & 64.96 & 82.48 & 2.28\\
    \hline
\end{tabular}
}
\raggedright \tiny $^{+}$ Multi-word prediction events occurred where more than one word was selected with one keystroke, as some suggested options by ChatBCI had more than one words. This occurred once for Subject S01 (\MWP{``AND-LOYAL''}) and once for Subject S05 (\MWP{``AND-THE-FUTURE''}). As such, for these subjects, where the number of keystrokes were equal or smaller than the total number of words in the sentence, KS has achieved or surpassed the theoretical KS-WP$_{\max}$ limit, resulting in the KS-DR of 0 and -1.37, respectively.\\
\raggedright \tiny $^{a}$ Numbers were rounded up using the ceiling function.
\end{threeparttable}
 \end{table*}
 
 \subsubsection{\underline{Keystroke Savings Analysis:}}
Table \ref{tab:5} presents results from keystroke analysis for Task 2 where subjects used ChatBCI to write improvised sentences. For each improvised sentence, the number of keystrokes used to write the sentence, the number of words, the number of occurrences of word completion (WCs) and word prediction (WPs) events along with KS, theoretical KS-WC$_{\max}$ and KS-WP$_{\max}$, and KS-DR are provided. Averages across subjects are also reported for each metric.

The averaged achieved keystroke savings was $80.68\%$, indicating that ChatBCI effectively reduced the number of keystrokes required to compose sentences. Indeed, for all subjects, KS exceeded the theoretical KS-WC$_{\max}$ limit and approached or exceeded the theoretical KS-WP$_{\max}$ limit, resulting in an average KS-DR of less than $2.50$\%.  Interestingly, for Subjects S01 and S05, there were instances where the suggestions made by GPT consisted of more than one word, allowing for the selection of multiple words in a single attempt. As such, the calculated KS for these two subjects reached or surpassed the theoretical KS-WP$_{\max}$ limit, since the number of used keystrokes were equal or smaller than the number of words in the improvised sentences. This resulted in KS-DR values of zero and negative, suggesting that ChatBCI can even provide greater efficiency than theoretically anticipated by maximizing the capabilities of LLMs.

%%%%%%%%%%%%%%%%%%%%%%%%%%%
%%%%%%%%%%%%%%%%%%%%%%%%%%%
%%%% Discussions
%%%%%%%%%%%%%%%%%%%%%%%%%%%
%%%%%%%%%%%%%%%%%%%%%%%%%%%

\section{Discussions}\label{discuss}
To the best of our knowledge, ChatBCI is the first P300 speller BCI to effectively and efficiently incorporate LLMs (here GPT), and leverage their predictive power and contextual understanding to enhance the speller's speed, performance, and overall user experience. Additionally, by offering three typing options (word prediction, word completion, and letter-by-letter), the new GUI in ChatBCI allows users to employ the speller based on their personal preferences when composing and typing sentences. As demonstrated in Section \ref{results}, compared to a traditional letter-by-letter P300 speller, ChatBCI offers significantly improved performance, including faster communication speeds, as evidenced by significantly shorter time to complete writing sentences, higher selection accuracy, and an significantly improved information transfer rate. These features reduce user fatigue and enhance user experience, making ChatBCI a desirable assistive technology candidate for individuals with motor and communication disabilities.

Another contribution of this paper is the introduction of keystroke analysis to assess the predictive typing capability of speller BCIs. Typically, the performance of P300 speller BCIs is assessed using metrics such as ITR, selection accuracy, and success rate. However, the integration of LLMs into speller BCIs introduces new capabilities, necessitating the development of additional metrics for performance evaluation. Here, we proposed to employ keystroke analysis to specifically assess the predictive typing capability of speller BCIs equipped with word prediction functionality. We also introduced a new metric, the keystroke savings deficit ratio (KS-DR), to evaluate how far a speller BCI with predictive features is from achieving its theoretical maximum keystroke savings limit. This metric can set targets for designing efficient speller BCIs. Moreover, the predictive typing features and the new keyboard GUI, which includes word-level suggestions as selectable keys, necessitate revisiting the usual way of calculating the information transfer rate. We introduced two approaches, one to estimate the lower bound (ITR$^*$-1), and another, as the observed measure of information transfer rate (ITR$^*$-2).

Additionally, we evaluated the performance of ChatBCI using a novel experiment of asking subjects to improvise sentences (online Task 2). Traditionally, speller BCIs have been evaluated by asking users to copy-spell pre-determined sentences. However, by having subjects improvise, we were able to assess the speller's real-world applicability, and effectiveness in handling spontaneous, user-generated content, resembling a more naturalistic conditions. Furthermore, compared to the copy-spelling task, improvisation may better resemble the need in most daily communications, since communication can be equivalently sufficient as long as the same intention, thought or request is conveyed, even using different words or sentences.

Table~\ref{tab:6} compares the performance of the proposed ChatBCI with recent P300 speller BCIs incorporating some level of word predictive typing functionalities in online spelling tasks. Performances are compared based on time to complete (min), selection accuracy (\%), success rate (\%), and ITR$^{*}$. For each metric, results averaged across subjects in the study are reported. Some studies did not report some of these metrics directly in their work. Additionally, since prior studies have not used keystroke savings analysis, we cannot make comparisons in this area. Note that due to variations in GUI design and experimental setups across studies, making an entirely fair comparison across studies is not possible. Additionally, \cite{speier2018improving} and \cite{chandravadia2022comparing} employed extra optimization features.

\begin{table*}
    \centering
        \caption{\small Performance comparison of ChatBCI with recent P300 speller BCIs incorporating some degrees of word predictive typing functionalities for online spelling tasks.}
    \label{tab:6}
    \begin{threeparttable}
    \resizebox{\textwidth}{!}{
\begin{tabular}{c|c|c|c|c|c|c|c|c|c}
    \hline
    \textbf{} & \textbf{NLP} & \textbf{\# of Words} & \textbf{User}& \textbf{Corrections} & \textbf{Optimization}& \textbf{Time to Complete (min)} & \textbf{Selection} & \textbf{SR} & \textbf{ITR$^*$} \\
    
        \textbf{Reference} & \textbf{} & \textbf{} &\textbf{Composed}& \textbf{Allowed} &  \textbf{Features}& \textbf{} & \textbf{Accuracy} & \textbf{} & \textbf{} \\
    
    \textbf{} &  \textbf{Technique}&\textbf{(\# of Char.)} & \textbf{(Y/N)} & \textbf{(Y/N)} & \textbf{(Y/N)} & \textbf{(Typing Speed (characters/min))} & \textbf{(\%)} & \textbf{(\%)}  & \textbf{(bits/min)} \\  
    \hline
    \cite{speier2018improving} & Probabilistic Automata & 10 (-) & N & N & Y &  N/A (N/A) & - & 94.80 & 59.39\tnote{b}\\
    
    \cite{chandravadia2022comparing} & Probabilistic Automata & 3 (30) & N & N & Y &  N/A (N/A) & - & 97.58 & 72.11\tnote{b} \\
  
     \cite{kaufmann2012spelling} & Dictionary & 9 (45) & N & Y & N & 12.40 (3.63)\tnote{e}  & - & 100 & 20.60 \\ 
   
     \cite{ryan2010predictive} & WordQ2 & 10 (58) & N & Y & N & 12.43 (5.28)\tnote{f} & 84.88  & 100  & -\\
   
     \cite{Akram2013novel} & Dictionary & 10 (-) & N & N & N & 16.60\tnote{a} (N/A) & 77.50 & - & - \\
     \hline
    \textbf{ChatBCI-Task 1(Control)} & None (LL) & 7\tnote{c} (26)\tnote{c} & Y (Offline) & Y & N & 27.55 (0.97)& 77.89 & 94.64& 6.72 \\
    
     \textbf{ChatBCI-Task 1} & GPT-3.5 & 7\tnote{c} (26)\tnote{c} & Y (Offline) & Y & N &10.43 (2.76) & 88.77 & 100 & 20.09  (26.28)\tnote{d} \\
     
     \textbf{ChatBCI-Task 2} & GPT-3.5 & 11\tnote{c} (62)\tnote{c}  & Y (Online) & N & N & \textbf{7.13} (\textbf{8.53})  & \textbf{93.16} & N/A & 64.76 (\textbf{83.46})\tnote{d} \\
     \hline
\end{tabular}
   }
\raggedright \tiny $^{a}$ This number was not directly reported in the paper, and is calculated here using the information provided ($1.66$ minutes per word, and a total of $10$ words). \\
\raggedright \tiny $^{b}$ ITR$^*$ is computed based on SR not selection accuracy.\\
\raggedright \tiny $^{c}$ Average across subjects rounded up using the ceiling function.\\
\raggedright \tiny $^{d}$ ITR$^*$-2 \\
\raggedright \tiny $^{e}$ Calculated as 45 characters divided by the reported average time to complete.\\
\raggedright \tiny $^{f}$ Calculated based on reported predictive output characters per minute.\\
\end{threeparttable}
\end{table*}

All the prior work listed in this table provided users with specific pre-determined words or sentences for copy-spelling tasks. In contrast, our work considered user-composed sentences in Task 1, and in Task 2, subjects composed sentences online and in real-time using ChatBCI. Therefore, our ChatBCI has been evaluated in an experimental setting that more closely resembles the real-world application of a speller BCI.

In terms of the speed, ChatBCI in Task 2, by taking full advantage of WC and WP features, achieves the shortest task completion time: an average of $7.13$ minutes for typing an average of $62$ characters.  In Task 2, ChatBCI achieved an average of $8.53$ characters/min across subjects for the typing speed representing an $61.55$\% improvement over the best previously reported experimental result for non-optimized BCI spellers \cite{ryan2010predictive} ($5.28$ characters per minute). As for selection accuracy and success rate, ChatBCI offers the best results in comparison to prior work, suggesting that the use of predictive features in the speller, can aid users to have less errors and mistakes. 

In terms of information transfer rate (ITR), the best results among prior work are reported by \cite{speier2018improving} and \cite{chandravadia2022comparing}. However, it is worth noting that unlike ChatBCI and other work, these two studies incorporated additional optimization features to enhance ITR$^*$ in their speller systems. In \cite{speier2018improving}, a dynamic stopping algorithm was used to reduce the number of flashes based on a threshold, while \cite{chandravadia2022comparing} employed different paradigms, with the Row-Column paradigm achieving an average ITR$^*$ of $72.11$ bits/min across subjects.  Comparatively, ChatBCI in Task 2, without incorporating optimization techniques, achieves an average lower bound ITR$^*$ (ITR$^*$-1) of $64.76$ bits/min, outperforming \cite{speier2018improving} and comparable to \cite{chandravadia2022comparing}. If we consider the observed ITR$^*$ (ITR$^*$-2),  ChatBCI in Task 2 offers the best average result, reaching $83.46$ bits/min. Looking at Table \ref{tab:4} and subject-level results, Subjects S06 achieves an impressive ITR$^*$-1 and ITR$^*$-2 values of 
$81.43$ bits/min and $107.49$ bits/min, respectively, for writing a sentence consisting of 104 characters. The next best ITR$^*$ values belong to Subjects S01 and S05, who had instances of selecting multiple words in a single selection.

Overall, by incorporating LLMs, ChatBCI represents a new generation of P300 speller BCIs, offering significant advancements in spelling efficiency, typing speed,  user experience, and overall performance, compared to existing spellers. Future work includes incorporating optimization techniques such as dynamic stop or inter-stimulus optimization, to further improve the performance.  
%%%%%%%%%%%%%%%%%%%%%%%%%%%
%%%%%%%%%%%%%%%%%%%%%%%%%%%
%%%% Conclusion
%%%%%%%%%%%%%%%%%%%%%%%%%%%
%%%%%%%%%%%%%%%%%%%%%%%%%%%
\section{Conclusion} \label{conc}
In this paper, we introduced ChatBCI, the first demonstration of a P300 speller BCI that incorporates LLMs (here GPT-3.5) to improve typing speed, performance, and user experience by offering predictive typing features. By enabling word and multi-word predictions through the integration of LLMs, ChatBCI reduces the need for letter-by-letter selection, cutting down on keystrokes, task time, and cognitive load.  ChatBCI also features a new keyboard GUI that offers users three typing options (word prediction, word completion, and letter-by-letter) allowing them to use the speller according to their personal preferences when composing sentences. To assess ChatBCI’s real-world applicability, we evaluated it not only through a traditional copy-spelling task but also through a new task, where users improvised sentences. This approach assesses speller's efficiency in practical, spontaneous communication scenarios, and its potential for real-time communication. Additionally, we introduced keystroke analysis and a new metric, the keystroke savings deficit ratio (KS-DR), to quantify speller's predictive typing performance. Results from two online tasks demonstrate that ChatBCI significantly outperforms traditional letter-by-letter P300 spellers. In the composition task, ChatBCI achieved $80.68\%$ keystroke savings and a new experimental record of $8.53$ characters per minute, averaged across subjects. With its multi-word prediction feature, ChatBCI can surpass the theoretical keystroke savings limit, underscoring the potential of LLM-integrated speller BCIs for practical use. ChatBCI represents the next generation of P300 speller BCIs, paving the way for real-time, practical solutions for individuals with motor and communication impairments.

\newpage
\section*{Appendix}
\begin{algorithm}
    \footnotesize
        \caption{\footnotesize Stepwise Feature Selection for SWLDA Classifier}\label{alg:1}
    \begin{algorithmic}[1]
        \State \textbf{Input:} EEG feature vector $\mathbf{f}=\{f_1,\cdots,f_i,\cdots,f_I\}$, class labels $\mathbf{y}\in\{0,1\}$ (presence or absence of the P300 evoked potential)
        \State \textbf{Output:} Selected feature subset $\mathbf{x}$ of $\mathbf{f}$ for LDA classification
        \State Initialize $\mathbf{x} = \emptyset$
        \State \textit{converged} $\gets$ False
        \While{not \textit{converged}}
            \State \textit{updated} $\gets$ False
            \State $\textbf{f}_{\mathrm{add}} \gets \{f_i|f_i\in \textbf{f},  f_i \notin \mathbf{x}\}$
            \For{each $f_i \in \textbf{f}_{\mathrm{add}}$}
            \State Acquire multi-variate OLS regression model $\mathbb{M}$, using $\mathbf{x} \cup \{f_i\}$ as independent variables, and the binary label $\mathbf{y}$ as dependent variable
            \State Evaluate $p_i$, the significance (\textit{p}-value) of $f_i$, in $\mathbb{M}$
           \EndFor
           \State Find $f_k \in \textbf{f}_{\mathrm{add}}$ with minimal $p$-value $p_k$
            \If{$p_k < 0.1$}
                \State Add $f_k$ to $\mathbf{x}$
                \State \textit{updated} $\gets$ True
            \EndIf
            \State Acquire multi-variate OLS regression model $\mathbb{M}$, using $\mathbf{x}$ as independent variables, and the binary label $\mathbf{y}$ as dependent variable
            \For{each $f_i \in \textbf{x}$}
                \State Evaluate $p_i$, the significance (\textit{p}-value) of $f_i$, in $\mathbb{M}$
            \EndFor 
            \State Find $f_k \in \textbf{x}$ with maximal $p$-value $p_k$
            \If{$p_k > 0.25$}
            \State Remove $f_k$ from $\mathbf{x}$
            \State \textit{updated} $\gets$ True
            \EndIf
            \If{not \textit{updated}}
                \State \textit{converged} $\gets$ True
            \EndIf
        \EndWhile
        \State \textbf{return} $\mathbf{x}$
    \end{algorithmic}
    \end{algorithm}

\newpage
\normalsize
\section*{References}
\bibliographystyle{IEEEtran}
\bibliography{refs}

\end{document}